\documentclass[aip,pop,preprint]{revtex4-2}
\usepackage{amsmath}
\usepackage{amssymb}
\usepackage{graphicx}
\usepackage{hyperref}
\graphicspath{{figures/}} % Location of the graphics files
\usepackage[usenames, dvipsnames]{color}

\newcommand{\Rm}{Rm}
\renewcommand{\Re}{Re}
\newcommand{\Ma}{M_{\mathrm{A}}}
\newcommand{\Pm}{Pm}
\newcommand{\rotp}{\mathbin{\rotatebox[origin=c]{21}{$'$}}}

\newcommand{\BT}{\textcolor{black}}

\begin{document}
\author{B.~Tripathi$^1$}
\email{btripathi@wisc.edu}
\author{A.E.~Fraser$^2$}
\author{P.W.~Terry$^1$}
\author{E.G.~Zweibel$^1$}
\author{M.J.~Pueschel$^{3,4}$}
\affiliation{
$^1$University of Wisconsin-Madison, Madison, Wisconsin 53706, U.S.A.\\
$^2$University of California, Santa Cruz, Santa Cruz, California 95064, U.S.A.\\
$^3$Dutch Institute for Fundamental Energy Research, 5612 AJ Eindhoven, The Netherlands\\
$^4$Eindhoven University of Technology, 5600 MB Eindhoven, The Netherlands
}

\title{Near-cancellation of up- and down-gradient momentum transport in forced magnetized shear-flow turbulence} 
%\title{Unmasking coherent vorticies: cancellation of up- and down-gradient momentum transport in magnetized shear flow turbulence} 

\today

\begin{abstract}

Visco-resistive magnetohydrodynamic turbulence, driven by a two-dimensional unstable shear layer that is maintained by an imposed body force, is examined by decomposing it into dissipationless linear eigenmodes of the initial profiles. The down-gradient momentum flux, as expected, originates from the large-scale instability. However, continual up-gradient momentum transport by large-scale linearly stable but nonlinearly excited eigenmodes is identified, and found to nearly cancel the down-gradient transport by unstable modes. The stable modes effectuate this by depleting the large-scale turbulent fluctuations via energy transfer to the mean flow. This establishes a physical mechanism underlying the long-known observation that coherent vortices formed from nonlinear saturation of the instability reduce turbulent transport and fluctuations, as such vortices are composed of both the stable and unstable modes, which are nearly equal in their amplitudes. The impact of magnetic fields on the nonlinearly excited stable modes is then quantified. Even when imposing a strong magnetic field that almost completely suppresses the instability, the up-gradient transport by the stable modes is at least two-thirds of the down-gradient transport by the unstable modes, whereas for weaker fields, this fraction reaches up to $98\%$. These effects are persistent with variations in magnetic Prandtl number and forcing strength. Finally, continuum modes are shown to be energetically less important, but essential for capturing the magnetic fluctuations and Maxwell stress. A simple analytical scaling law is derived for their saturated turbulent amplitudes. It predicts the fall-off rate as the inverse of the Fourier wavenumber, a property which is confirmed in numerical simulations.

\end{abstract}
\maketitle

%%%%%%%%%%%%%%%%%%%%%%%%%%%%%%%%%%%%%%%%%%%%%%%%%%%%%%%%%%%%%%%%%%%%%%%%%%%%%%%%%%%%%%%%%%%%%%%%%%%%%
%%%%%%%%%%%%%%%%%%%%%%%%%%%%%%%%%%%%%%%%%%%%%%%%%%%%%%%%%%%%%%%%%%%%%%%%%%%%%%%%%%%%%%%%%%%%%%%%%%%%%
%%%%%%%%%%%%%%%%%%%%%%%%%%%%%%%%%%%%%%%%%%%%%%%%%%%%%%%%%%%%%%%%%%%%%%%%%%%%%%%%%%%%%%%%%%%%%%%%%%%%%
\section{Introduction}
\label{sec:intro}

Owing to their ubiquity in laboratory,\cite{harding2009} geophysical,\cite{hasegawa2004, waugh2017} and astrophysical environments,\cite{ read2020, fuller2019, pessah2006, goodman1994, alfves2020, fleck2020} shear layers have been extensively studied. \cite{miura1999, lecoanet2016} Observations and analyses from experiments and direct numerical simulations have offered insights into the connection between large-scale vortical structures formed from the instability of a shear layer and turbulent transport across the layer.\cite{ho1984, browand1983, starr1970} Properties like shape and scale of the nonlinearly saturated vortices, which dominate the transport, are generally attributed to the linearly-unstable eigenmodes or closely related nonlinear fluctuations. \cite{miura1978, horton1987} The nonlinear saturation of the instability, however, can be more complex than just the finite-amplitude modifications of unstable modes, as emerging understanding in fusion plasma instability demonstrates.\cite{terry2021, whelan2018, pueschel2016, makwana2014, hatch2013, hatch2011, pueschel2021, li2022, whelan2019, terry2018, fraser2018, makwana2011, makwana2012,  terry2006}

\BT{Already in the late 1960s, using one of the early numerical simulations of shear instability,\cite{levy1968} it was hinted that the nonlinear saturation of Kelvin-Helmholtz instability involves, contrary to finite-amplitude modifications of unstable modes, quasi-periodic oscillations in the fluctuations. Later, an intuitive understanding of how such a phenomenon occurs in sheared fluids\cite{zabusky1971} has been reported by invoking vortex nutation:\cite{miura1978} Fluctuation-amplitude oscillations correlate with oscillations in the mean flow energy and lead to vortex nutation. Fluctuations, however, are usually not decomposed into the complete set of linear eigenmodes, and are commonly assumed\cite{miura1978} to be due to unstable mode structures. But since unstable modes always drive a down-gradient momentum transport, they cannot explain the increase of kinetic energy in the mean flow.} 

\BT{Notably, occasional up-gradient momentum transport has been observed in several experimental and numerical studies where an unstable shear layer drives the turbulence.\cite{huang1990,moser1993,riley1980, oster1982, ito2013} Analyses of these transient events \cite{vandine2021, lopez2018, ho1984, hussain1985, hussain1986} do not address the underlying conditions producing this dynamics---whether the transient up-gradient transport is a part of an ongoing subdominant process with occasional breakthroughs, or simply spontaneous fluctuations. The laboratory and prior numerical experiments alone are not sufficient to definitively answer this question. One way to expose the underlying process is to examine the turbulent fluctuations using a complete eigenmode decomposition, and assign roles and activities to each mode in the transport phenomena. Indeed there can be modes other than the unstable modes that are important in the turbulent phase, as an insightful study hints: the dominant vortex in a turbulent background orients quasi-periodically against (or towards) the mean flow and drives the down-gradient (or up-gradient) momentum transport.\cite{ho1984} To understand such behaviors in detail, it is instructive to also analyse how the instability saturates, a question that has long been of interest\cite{landau1944} but for which understanding remains incomplete.}

When turbulence is sustained via continuous energy injection from a large-scale instability, there exist two primary candidates for instability saturation. A common (but not necessarily justified) assumption is that energy injected by the instability is transferred conservatively to increasingly smaller scales in a forward, Kolmogorov-like cascade, where nonlinear interactions move energy between linearly unstable or marginal modes until a dissipation range is reached at small scales.\cite{fuller2019} An alternative process involves linearly stable eigenmodes at the large injection scales, which absorb and remove significant energy from scales that launch the inertial cascade. In several studies of microturbulence in fusion plasmas, linearly stable modes have been found to be excited to significant levels via nonlinear interactions and to drastically affect the saturated amplitudes and transport characteristics of the system.\cite{terry2021, whelan2018, terry2018, fraser2018, pueschel2016, makwana2014, hatch2013, makwana2012, hatch2011, makwana2011, terry2006} 

Stable modes in shear flow turbulence, however, have been studied only recently\cite{fraser2017, fraser2021, tripathi2022} and more remains to be understood, e.g., their role in mixing and magnetic field evolution and how they might affect reduced models of turbulence and transport. It was predicted in Ref.~\cite{fraser2017} that the Kelvin-Helmholtz instability in its nonlinear evolution excites a linearly-stable conjugate-root\cite{chandrashekhar1961} of the inviscid instability, which affects the instability saturation even when viscosity is finite. This was later verified in numerical simulations of freely evolving shear layers. \cite{fraser2021} However, the rapid relaxation of the layer towards a stable profile on a time scale similar to that of stable-mode excitation prevented general conclusions from being reached, regarding how the turbulence and transport are affected by the stable modes. The issue is aggravated by the addition of a flow-aligned magnetic field, which causes the layer to relax even more rapidly. To circumvent this challenge, one may drive the mean flow towards the unstable profile and thus achieve a thorough statistical quantification of the stable modes. Note that driven profiles are quite common in astrophysical shear flows, with forces like gravity providing free energy for the drive. For these reasons, driven shear flow is studied here.

The principal result of this study is that significant up-gradient momentum transport is driven by nonlinearly excited (linearly-)stable modes, cancelling a substantial portion of the down-gradient transport by unstable modes, and notably this transport is present not just during turbulent momentum flux reversals, but is continuously at work at a slightly lower level than that of the unstable modes. This finding is robust even for variations of orders of magnitude in background magnetic field strength, magnetic Prandtl number (or resistivity), and forcing strength of the mean flow. Note that the stronger background magnetic field tends to suppress the instability \cite{chandrashekhar1961} and disrupt the large-scale vortices, \cite{mak2017} while larger magnetic Prandtl number (weaker resistivity for a fixed viscosity) extends the scale range of magnetic fluctuations, compared to the flow fluctuations. \cite{schekochihin2002} We also show, for astrophysical applications, that a turbulent viscosity can be defined, with the addition of stable modes, that can reliably capture the Reynolds stress: Without stable modes, however, the stresses are greatly over-predicted by the unstable modes.

This article is organized in the following manner. Section~\ref{sec:sec2} entails the magnetohydrodynamic (MHD) model of the shear flow and details the system set-up. In Sec.~\ref{sec:sec3}, the complete linear eigenspectrum is presented, along with a discussion on the roles of different eigenmodes. Section~\ref{sec:sec4} shows the full nonlinear evolution of MHD Kelvin-Helmholtz instability using direct numerical simulations. A decomposition of the turbulent fluctuations onto linear eigenmodes is performed in Sec.~\ref{sec:sec5}, where a detailed study of imprints of stable modes in turbulence and transport is presented. Section~\ref{sec:sec6} summarizes the findings of this work.

\section{Model and simulation set-up} \label{sec:sec2}
An incompressible magneto-fluid is considered in this study, and standard MHD equations are adopted:
\begin{subequations}
\begin{align}
    &\mathbf{\nabla} \cdot \mathbf{u}=0,\\  \label{eq:mhdu}
    &\partial_t \mathbf{u} + \mathbf{u} \cdot \mathbf{\nabla} \mathbf{u} = -\frac{\nabla P}{\rho} + \frac{\left(\nabla \times \mathbf{B}\right) \times \mathbf{B}}{4\pi \rho} + \nu \nabla^2 \mathbf{u} + \mathbf{f},\\
    &\mathbf{\nabla} \cdot \mathbf{B}=0,\\  \label{eq:mhdb}
    &\partial_t  \mathbf{B} = \nabla \times \left(\mathbf{u} \times \mathbf{B} \right) + \eta \nabla^2 \mathbf{B},
\end{align}
\end{subequations}
where $\mathbf{u}$, $\mathbf{B}$, $P$, $\rho$, $\nu$, $\eta$, and $\mathbf{f}$ respectively denote the fluid velocity, magnetic field, pressure, fluid density, viscosity, ohmic diffusivity, and externally supplied acceleration to the magneto-fluid. 
%We wish to deal with a $2$D system for the following reasons: (i) the physics to be investigated in this paper is relatively new and very little is known about the stable eigenmodes in quasi-stationary state of MHD turbulence; (ii) such novelty is better explored with comparatively less computational resources in $2$D. Despite this, the computational hours needed for some of the simulations in this paper are formidable (e.g., our simulation with magnetic Prandtl number, $\Pm$, of $10$). 

\subsection{Background flow, magnetic field, and forcing}
A shear layer is examined on a two-dimensional $(x,z)$ plane with the initial fluid velocity given by $\mathbf{u}(x,z,t=0) = U_0 \mathrm{tanh}(z/a) \hat{\mathbf{x}}$ and a flow-aligned magnetic field, initially uniform, as $\mathbf{B}(x,z,t=0) = B_0 \hat{\mathbf{x}} $. The parameters $a, U_0,\mathrm{ and\ } B_0$ signify the half-width of the flow-shear, maximum initial fluid velocity, and initial magnetic field, respectively. These parameters are exploited to non-dimensionalize all the variables henceforth. Length, time, and energy (per unit mass) are hereafter measured in units of $a$, $a/U_0$, and $U_0^2$, respectively. Thus the initial (or reference) flow and magnetic field are represented by $U_\mathrm{ref}(z)= \mathrm{tanh}(z)$ and $B_\mathrm{ref}(z)=1$ in the rest of this article. The ratio of the maximum fluid speed $U_0$ to the Alfvén speed can be written as the Alfvénic Mach number $\Ma=U_0 \sqrt{4\pi \rho}/B_0$. The viscosity and resistivity are quantified via fluid Reynolds number $\Re=U_0 a/\nu$ and magnetic Reynolds number $\Rm=U_0 a/\eta$, respectively. 

In two dimensions, a more convenient formalism is available, using the streamfunction $\phi$ and flux function $\psi$. Defining $\mathbf{u}= \mathbf{\hat{{y}}} \times \nabla \phi$ and $\mathbf{B}= \mathbf{\hat{{y}}} \times \nabla \psi$, the vorticity and the current become $\nabla^2 \phi \mathbf{\hat{{y}}}$ and $ \nabla^2 \psi \mathbf{\hat{{y}}}$, respectively. Taking the curl of Eq.~\eqref{eq:mhdu}, and rewriting Eq.~\eqref{eq:mhdb} in terms of the stream- and flux-functions yields \cite{biskamp2003}
\begin{subequations}
\begin{align} \label{eq:momentumeqn}
    \begin{split}
    &\partial_t \nabla^2 \phi + \{\nabla^2 \phi , \phi \} = \Ma^{-2} \{\nabla^2 \psi , \psi \} + \Re^{-1} \nabla^4 \phi +   \partial_z f(k_x\mathrm{=}0, z, t),
    \end{split}\\ \label{eq:inductioneqn}
    &\partial_t  \psi = \{ \phi , \psi \} + \Rm^{-1} \nabla^2 \psi,
\end{align}
\end{subequations}
where the Poisson bracket is $\{P, Q\}=\partial_x P \cdot \partial_z Q - \partial_z P \cdot \partial_x Q$; e.g., $ \{ \phi , \psi \} = -\mathbf{u} \cdot \mathbf{\nabla} \psi $. Here, $k_x$ is the Fourier wavenumber along the $x$-axis. The parameters $\Re=\Rm=500$ are chosen for all simulations unless mentioned otherwise (where $\Rm$ is changed to $50$ and $5\,000$ in different simulations). It should be emphasized that these Reynolds numbers are defined using the initial scale $a$ of the sharpest gradient in the flow as the characteristic length scale; however, as the system evolves nonlinearly via vortex merging, despite the forced mean flow, eddies of the size of the simulation box appear, which may be considered as the characteristic length scale of motion.\cite{lecoanet2016} When choosing this normalization, non-dimensional numbers should be scaled accordingly, e.g., $\Rm=5\,000$ becomes $\Rm=5\,000\times L_x \approx 1.5 \times 10^{5}$, where $L_x$ represents the box-size along the mean flow direction. The external body force, $\mathbf{f} = f(k_x\mathrm{=}0, z, t) \mathbf{\hat{x}}$, is applied to the mean flow only, which is highlighted in Eq.~\ref{eq:momentumeqn} using the explicit mention of $k_x\mathrm{=}0$. As in a recent study,\cite{tripathi2022} the forcing drives the instantaneous mean flow towards the initial unstable profile $U_\mathrm{ref}(z)$. A similar forcing mechanism exists for geo- and astrophysical flows where gravitation\cite{ebrahimi2009} tends to build shear layers. We assume here such a force, represented as a Krook operator, \cite{pueschel2014, marston2008, smith2021} as
\begin{equation} \label{eq:forcing}
     f = D_{\mathrm{Krook}} \left[U_{\mathrm{ref}}(z) - \langle U(x,z,t) \rangle_x\right] + F_0,
\end{equation}
where $D_\mathrm{Krook}$, sometimes also referred to as the profile relaxation rate,\cite{allawala2020} measures the forcing strength (in units of $U_0/a$);  and $\langle U(x,z,t) \rangle_x$ represents the instantaneous $x$-averaged flow. If $D_\mathrm{Krook}=0$, the shear layer evolves freely and decaying turbulence is realized as a result of the Kelvin-Helmholtz instability and the turbulence it generates. 
%This was studied exhaustively by Fraser \textit{et al.}\cite{fraser2021} where a reduced description of the turbulent flow was sought. Because of the rapid relaxation of the unstable profile towards the stable one, the eigenmodes of the system did not conform to a reduced description of the turbulent flow although a reduced description of the Reynolds stress across the middle of the shear layer was remarkably achieved. This progress hinted that a forced shear layer perhaps is capable of driving a quasi-stationary turbulence that is amenable to an extreme reduction in the representation of the turbulent flow as well. This is one of the goals of the current article.

The time-independent force $F_0$ is implemented only to balance the viscous diffusion of the initial shear layer: $\Re^{-1} \nabla^2 U_{\mathrm{ref}}(z)  + F_0 = 0$ ensures an initial equilibrium state to which small-amplitude perturbations are added before the system is evolved.

\subsection{Initial and boundary conditions}
As in the unforced study, \cite{fraser2021} a simulation box of $L_x=10\pi$ is considered, but double the size along the $z$-axis ($L_z=20\pi$), given that the quasi-stationary turbulence simulated herein is run for much longer time, which tends to create fully developed turbulent features that are larger in size. Thus we adopt a larger domain to minimize their potential interactions with the boundaries in the $z$-axis. Note the forcing applied to the mean flow prevents profile relaxation and the turbulence remains mostly in the vicinity of the shear layer. The numerical code Dedalus, \cite{burns2020} a pseudospectral solver, is used in this study. Fourier modes along the $x$-axis and Chebyshev polynomials along the $z$-axis are employed with $(N_x, N_z)=(2048, 2048)$ spectral modes. We confirmed that the spectral energy density and dissipation are converged at this resolution. Note also that these high resolutions benefit the eigenmode projection of nonlinear data in the post-processing analysis. Only for the simulation with magnetic Prandtl number of $10$, the box size was changed to $(L_x, L_z)=(6\pi, 8\pi)$ and the resolution was increased to $(N_x, N_z)=(4096, 4096)$; the same simulation was repeated with $(N_x, N_z)=(4096, 8192)$, but only for early times due to computational cost, and found to reproduce, among others, the energy evolution. All simulations use $3/2$ dealiasing rule, additionally. The boundary conditions used in all simulations are periodic along the $x$-axis; and perfectly conducting, no-slip, co-moving (with the initial flow) at the top and bottom boundaries, $z=\pm L_z/2$.\cite{fraser2021, tripathi2022}

The initial equilibrium state is seeded with small-amplitude perturbations $(\widetilde{\phi}, \widetilde{\psi})$ at all Fourier wavenumbers, as\cite{fraser2021}
\begin{equation}
\begin{aligned}[b]
    \widetilde{\phi}(x,z,t=0) 
    &= A_\phi \sum_{k_x \neq 0} k_x^a e^{i r_\phi(k_x)} e^{-z^2/\sigma^2} e^{ik_x x},
\end{aligned}
\end{equation}
and
\begin{equation}
    \widetilde{\psi}(x,z,t=0) = A_\psi \sum_{k_x \neq 0} k_x^a e^{i r_\psi(k_x)} e^{-z^2/\sigma^2} e^{ik_x x}.
\end{equation}
Here, $A_\phi$ and $A_\psi$ set the overall amplitudes of the perturbations that have a Gaussian width controlled by $\sigma$ and the rate at which they fall-off with the wavenumbers given by $a$. The random phases $r_\phi(k_x)$ and $r_\psi(k_x)$, forming a uniform distribution in $[0, 2\pi)$, are issued for each different $k_x$ using a pseudo-random number generator. Different choices of these initial conditions were investigated in Ref.~\cite{fraser2021}, motivating the choice here: $a=-1$, $\sigma=2$, and $A_\phi=A_\psi = 10^{-3} $. This set of parameters offers distinct linear and nonlinear phases of evolution.

%%%%%%%%%%%%%%%%%%%%%%%%%%%%%%%%%%%%%%%%%%%%%%%%%%%%%%%%%%%%%%%%%%%%%%%%%%%%%%%%%%%%%%%%%%%%%%%%%%%%%
%%%%%%%%%%%%%%%%%%%%%%%%%%%%%%%%%%%%%%%%%%%%%%%%%%%%%%%%%%%%%%%%%%%%%%%%%%%%%%%%%%%%%%%%%%%%%%%%%%%%%
\section{Linear eigenmodes} \label{sec:sec3}
Aiming to understand the nonlinear excitation of linear eigenmodes in the turbulent phase, first the nonlinear initial-value problem is solved to collect high-fidelity turbulent data. Afterward, a separate eigenvalue problem is solved to obtain a complete linear eigenspectrum and eigenmodes, which are used to expand the nonlinear data on this basis to track the amplitude of each eigenmode. Such a basis is obtained by linearizing the governing equations around the initial flow and magnetic field profiles, by dropping the dissipative terms. The eigenmodes thus obtained are of a dissipationless linear operator. Of course, the meaning and utility of this linear basis is \textit{a priori} unknown. Nevertheless, when a basis forms a \textit{complete} set, one can always expand an arbitrary fluctuation on that basis. As the non-dissipative equations of motion preserve Parity-Time (PT-)reversal symmetry, such a system is theoretically guaranteed to yield a complete basis as established recently in PT-symmetric quantum mechanics.\cite{bender2019} Previous studies in gyrokinetic and MHD plasmas have also revealed the usefulness of dissipationless linear eigenmodes in interpreting dissipative nonlinear systems. \cite{fraser2018, fraser2021, hatch2016}

\subsection{Complete eigenspectrum} \label{sec:sec3a}
With the intent of obtaining dissipationless linear eigenmodes, the variables $(\phi,\psi)$ in Eqs. \eqref{eq:momentumeqn} and \eqref{eq:inductioneqn} are decomposed into background and perturbations, $(\phi,\psi) = (\phi_\mathrm{ref},\psi_\mathrm{ref}) + (\widetilde{\phi},\widetilde{\psi})$. The linearized dissipationless equations for the evolution of perturbations are
\begin{subequations} {\label{eq:lineqns}}
    \begin{align} {\label{eq:lineqnsa}}
    \partial_t \nabla^2 \widetilde{\phi} &= -\left[U_\mathrm{ref} \partial_x \nabla^2 - (\partial_z^2 U_\mathrm{ref})\cdot \partial_x \right] \widetilde{\phi}+ \frac{1}{\Ma^2} \left[B_\mathrm{ref} \partial_x \nabla^2 - (\partial_z^2 B_\mathrm{ref})\cdot \partial_x \right] \widetilde{\psi},\\ \label{eq:lineqnsb}
    \partial_t \widetilde{\psi}  &=  -U_\mathrm{ref} \partial_x \widetilde{\psi} + B_\mathrm{ref} \partial_x \widetilde{\phi}.
    \end{align}
\end{subequations}
Fourier transforming along the $x$-axis and assuming time variation at each Fourier wavenumber takes the form $\hat{\phi}(k_x,z, \omega) e^{i\omega(k_x) t}$, Eqs.~\eqref{eq:lineqnsa}-\eqref{eq:lineqnsb} become
\begin{subequations} {\label{eq:lineqnsfft}}
    \begin{align} \label{eq:lineqnsffta}
    \omega \left(\partial_z^2-k_x^2 \right)\hat{\phi} &= - k_x \left[U_\mathrm{ref}  \left(\partial_z^2-k_x^2 \right) - (\partial_z^2 U_\mathrm{ref})  \right] \hat{\phi}+ \frac{1}{\Ma^2}  k_x \left[B_\mathrm{ref}  \left(\partial_z^2-k_x^2 \right) - (\partial_z^2 B_\mathrm{ref}) \right] \hat{\psi},\\ \label{eq:maginduction}
    \omega \hat{\psi}  &=  -U_\mathrm{ref} k_x  \hat{\psi} + B_\mathrm{ref} k_x  \hat{\phi}.
    \end{align}
\end{subequations}

\begin{figure*}[htbp!]
	\includegraphics[width=1\textwidth]{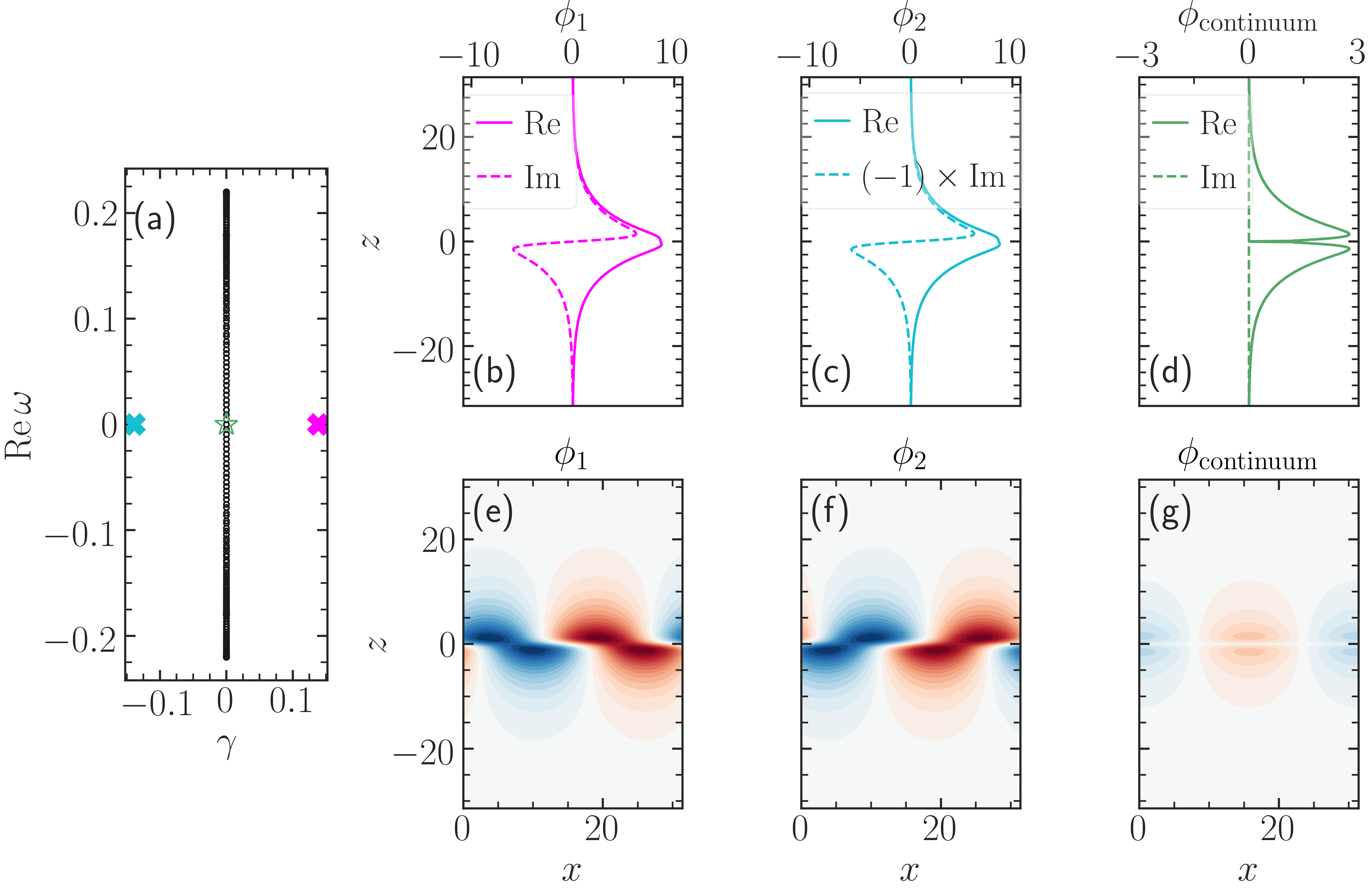}
	\caption{(a) Linear eigenspectrum of the MHD shear-flow system (at $k_x=0.2$ with $M_\mathrm{A} = 10$). The unstable and stable modes are shown with thick (colored) crosses. Among the continuum modes that form a vertical line, a zero-frequency continuum mode is displayed with a green-colored star.} (b)--(d) Eigenfunctions in $z$-space, with real (Re) and imaginary (Im) parts, for unstable ($\phi_1$), stable ($\phi_2$), and one continuum ($\omega = 0$) mode. (e)--(g) Corresponding eigenmode structures in ($x,z$) space. Note that the eigenmodes $\phi_1$ and $\phi_2$ are complex conjugate to each other. Imaginary parts in their eigenfunctions induce relative tilt between them in $(x,z)$ space, which will be consequential for momentum transport in Sec.~\ref{sec:competeunstablestable}. Each eigenmode is normalized to have unit total energy [following which the maximum values of $\phi$ in (b)--(g) are chosen].\label{fig:fig1}
\end{figure*}

Solving Eqs.~\eqref{eq:lineqnsffta}-\eqref{eq:maginduction}, the eigenvalues $\omega$ are found to be real except when $0 < |k_x| < 1$, where two of the real eigenvalues coalesce to produce imaginary eigenvalues, \cite{fu2020} as complex conjugate to each other. These are the growth rates of the unstable eigenmode and its conjugate stable eigenmode, which evolve in time as $\textrm{e}^{\gamma(k_x)t}$ and $\textrm{e}^{-\gamma(k_x)t}$, respectively. This mode-pair is shown, for the first Fourier wavenumber $k_x=2\pi/L_x=0.2$, in Fig.~\ref{fig:fig1}(a), along with all the purely real eigenvalues. The latter constitute the eigenmode continuum \cite{case1960} and are theoretically infinite in number, although numerical discretization always yields a finite but very large number of modes ($>3,000$ for each wavenumber in this study). These eigenvalues are given by the relation $\omega/k_x +U_\mathrm{ref}(z) \pm v_{\mathrm{A, ref}}(z)=0$, where $ v_{\mathrm{A, ref}}(z)$ is the Alfvén speed along the reference magnetic field at the vertical coordinate $z$.

The eigenfunctions, normalized to have unit total energy, are also shown in Fig.~\ref{fig:fig1}: along the $z$-axis, see Figs.~\ref{fig:fig1}(b)--(d), and in $(x,z)$ space, see Figs.~\ref{fig:fig1}(e)--(g). Note that complex conjugation transforms the unstable mode $\phi_1(k_x,z)$ into the stable mode $\phi_2(k_x,z)$ and vice-versa. This is a direct consequence of spontaneous PT-symmetry breaking in the ideal shear-flow instability. \cite{fu2020} (The spontaneous symmetry breaking does not imply that the equation of motion or the associated Hamiltonian breaks PT-symmetry; it is rather some of the eigenfunctions of such a PT-symmetry-preserving Hamiltonian that break PT-symmetry.)
%Nomenclature as such can be confusing to some erudite readers, but the standard conventions are followed to promote cross-disciplinary knowledge that can aid each other in the future.

A representative eigenfunction of a continuum mode, shown in Fig.~\ref{fig:fig1}(d), exhibits sharp and narrow structure. To what physics each type of eigenmode structure contributes will be explored in this article. 

\subsection{Competing roles of unstable and stable modes} \label{sec:competeunstablestable}
\begin{figure*}[htbp!]
	\includegraphics[width=1\textwidth]{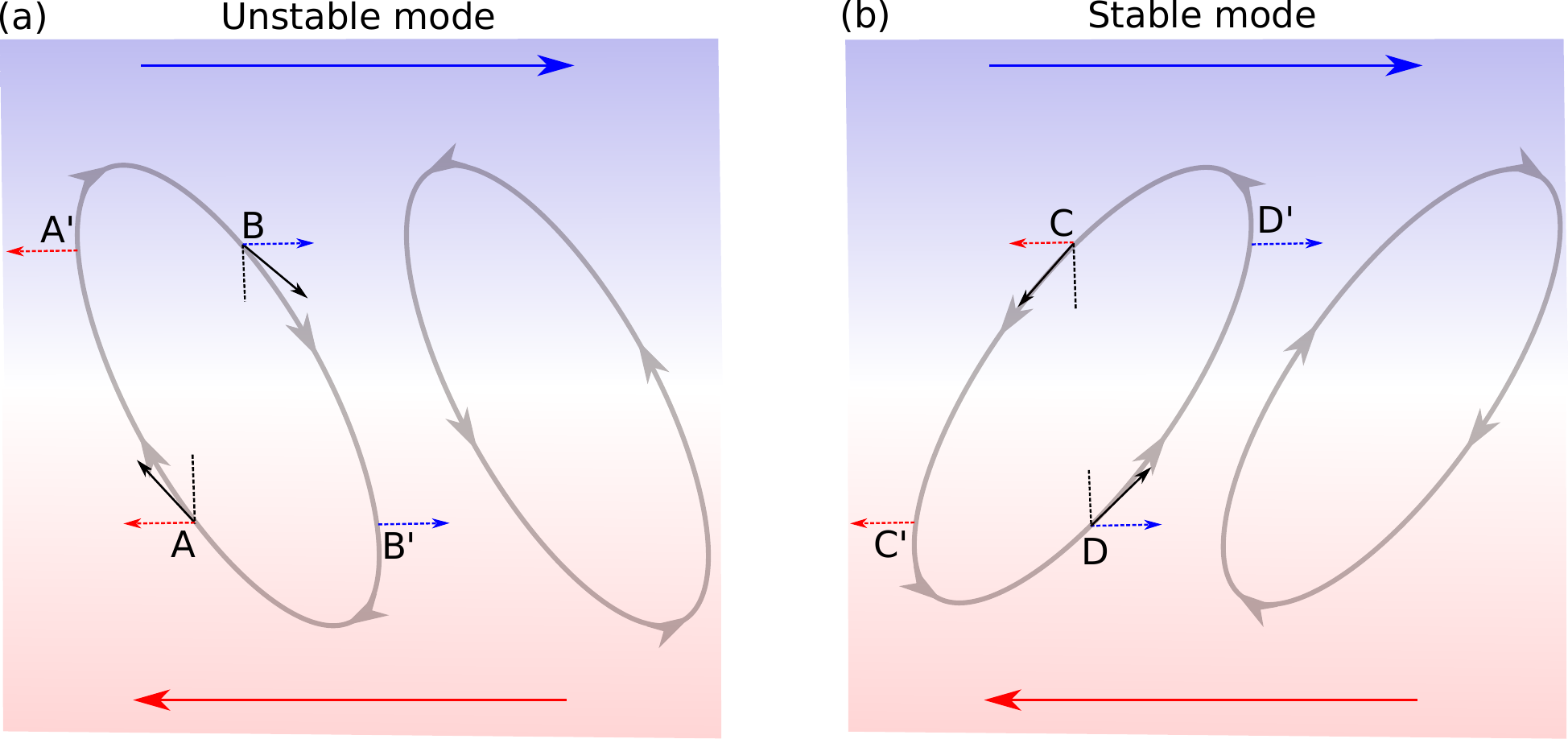}
	\caption{(a) The unstable mode of the flow transport momentum in the down-gradient direction: $-x$-directed momentum at A is carried to A$\rotp$ and $+x$-directed momentum at B is carried to B$\rotp$. Fluxes A$\to$A$\rotp$ and B$\to$B$\rotp$ act to relax the mean flow gradient (shown with the long horizontal arrows). (b) Oppositely tilted eddies, which correspond to a stable mode, transport momentum in the up-gradient direction: $-x$-directed momentum at C is carried to C$\rotp$ and $+x$-directed momentum at D is carried to D$\rotp$. Both of these fluxes replenish the mean flow. The direction of the streamlines (shown with grey arrows on the elliptic eddies) does not alter these properties, but the tilt does.}\label{fig:fig2}
\end{figure*}

Shown in Fig.~\ref{fig:fig2} is a schematic diagram, illustrating how the relative tilts in the eddies can transport momentum in opposite directions across the shear layer.\cite{starr1970} It can be qualitatively observed from Figs.~\ref{fig:fig1}(e) and (f) that the unstable and stable modes, drive down- and up-gradient momentum transport, respectively. Precise quantative measures will be built and computed later in Sec.~\ref{sec:transportmodeling}.

Since the unstable and stable modes compete with each other to transport momentum in opposing directions, the excitation levels of these modes are crucial. In the linear phase of instability evolution, the transport by the unstable modes dominates over the transport by the stable modes. However, this need not be the case in the nonlinear phase, as nonlinear processes can excite the stable modes to appreciable levels. Whenever the stable modes surpass the unstable modes in amplitudes, net momentum is transported in the up-gradient direction. \cite{terry2009, fraser2021} In extremely simplified models of transport, such as eddy viscosity models, this contributes to negative eddy viscosity. Computing the amplitude of each eigenmode in the nonlinear phase can thus be helpful to build improved reduced transport models. A recent investigation also demonstrated that this kind of competition between the two large-scale eigenmodes alters the magnetic cascade substantially. \cite{tripathi2022}

%%%%%%%%%%%%%%%%%%%%%%%%%%%%%%%%%%%%%%%%%%%%%%%%%%%%%%%%%%%%%%%%%%%%%%%%%%%%%%%%%%%%%%%%%%%%%%%%%%%%%
%%%%%%%%%%%%%%%%%%%%%%%%%%%%%%%%%%%%%%%%%%%%%%%%%%%%%%%%%%%%%%%%%%%%%%%%%%%%%%%%%%%%%%%%%%%%%%%%%%%%%
\section{Nonlinear evolution}\label{sec:sec4}
Having provided a description of linear eigenmodes, we now turn to properties of the nonlinear system, before discussing how expressions of linear modes may be identified in turbulent fluctuations.

\subsection{Finte-amplitude Kelvin-Helmholtz instability}
\begin{figure*}[htbp!]
	\includegraphics[width=0.85\textwidth]{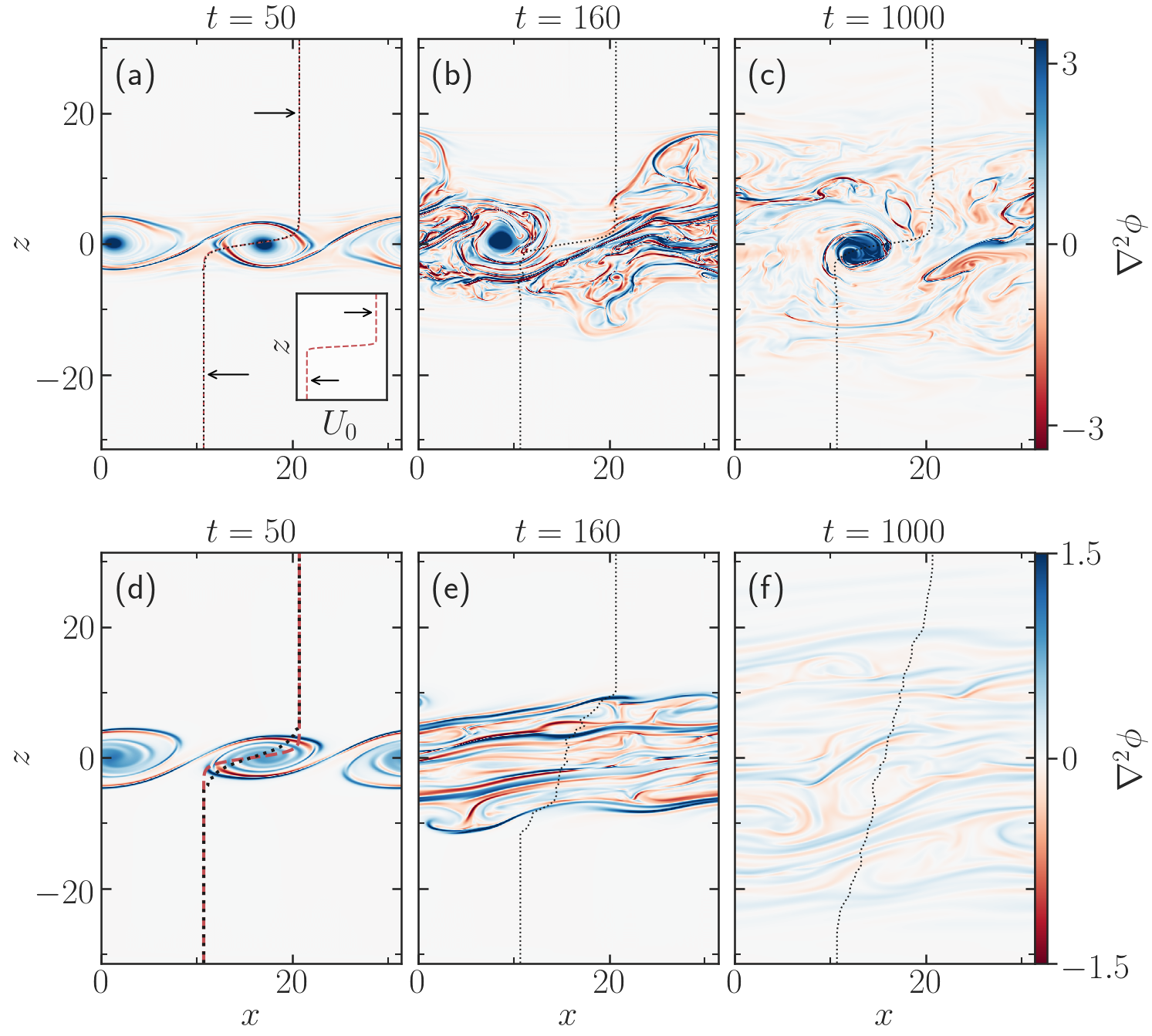}
	\caption{Time evolution of vorticity in (a)--(c) in a simulation with a forced background flow, $D_\mathrm{Krook}=2$; and (d)--(f) in a simulation with a freely evolving shear layer, $D_\mathrm{Krook}=0$; both for $M_\mathrm{A}=10$. Panels (a)--(c) share the same colorbar, and (d)--(f) share another colorbar. The instantaneous mean flow profile in each of the subplots is shown with a black dotted curve, where the vertical axis represents the $z$-coordinate and the horizontal direction corresponds to the $x$-velocity $U_0(z,t)$, as exemplified in the inset of (a). Two arrows pointing in opposite directions show the direction of the flow in the regions $z>0$ and $z<0$. The initial flow profile $U_0(z,t=0)=\mathrm{tanh}(z)$ is shown with a red dashed curve in (a) and (d). Rapid profile flattening is evident in (d). While the instability dies out in the unforced case, quasi-stationary turbulence is realized in the forced case in (c).}\label{fig:fig3}
\end{figure*}

Small-amplitude perturbations in the flow and magnetic field evolve exponentially fast in the linear regime of the instability, giving rise to a chain of spiral vortices, as evident in Figs.~\ref{fig:fig3}(a) and (d). These structures then interact nonlinearly with nearby vortices to yield even larger turbulent structures as in Figs.~\ref{fig:fig3}(b) and (c). A contrast is to be made between forced and unforced simulations. In the latter, the gradient of the mean flow flattens out as the instability extracts energy. Decaying turbulence then ensues. Forcing the mean flow, however, leads to a quasi-stationary turbulence, as the energy in the gradient is replenished with the instability drawing on its energy. In the saturated stage, energy input through the unstable modes is balanced by energy removal via stable modes as well as dissipative channels.

\subsection{Momentum transport}

\begin{figure*}[htbp!]
	\includegraphics[width=1\textwidth]{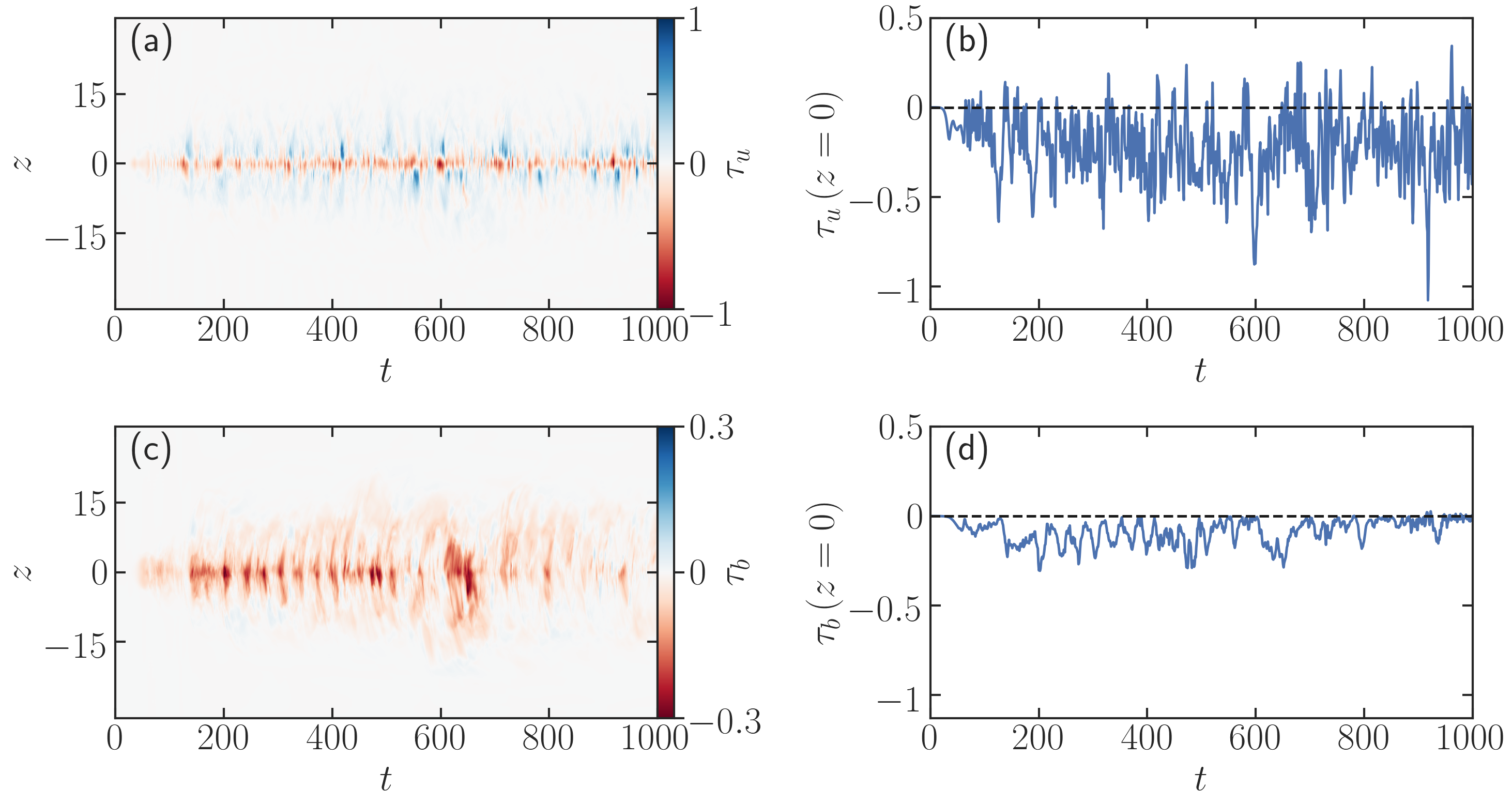}
	\caption{Time evolution of MHD stresses. (a) Reynolds stress $\tau_u(t,z)$. (b) Reynolds stress $\tau_u(t,z=0)$ at the middle of the layer, $z=0$. (c) Maxwell stress $\tau_b(z,t)$. (d) Maxwell stress $\tau_b(t,z=0)$ at $z=0$. All data shown are for a single simulation with $M_\mathrm{A}=10$, $D_\mathrm{Krook}=2$. The Reynolds stress in (b) reverses several times, in contrast to the Maxwell stress in (d), which is almost always down-gradient.}\label{fig:fig4}
\end{figure*}

It is now timely to discuss the turbulent transport of momentum in nonlinear simulations. To derive the turbulent stresses, the evolution equation of the mean flow can be written by $x$-averaging the momentum equation,
\begin{equation} \label{eq:stresseqn}
    \frac{\partial}{\partial t}\langle U \rangle_x = -\frac{\partial}{\partial z}\left(\tau_u +\tau_b\right) + D_\mathrm{Krook} \left[U_\mathrm{ref}(z)-\langle U \rangle_x \right] + \frac{1}{Re} \frac{\partial^2 }{\partial z^2} \left[\langle U \rangle_x - U_\mathrm{ref}(z) \right],
\end{equation}
where $U=U(x,z,t)$ represents the instantaneous flow, $\langle \cdot \rangle_x$ signifies $x$-averaging operation, and $\tau_u$ and $\tau_b$ are the Reynolds and Maxwell stresses, arising from the correlations of turbulent fluctuations of the flow and the magnetic fields, respectively. Note that in Fraser \textit{et al.}\cite{fraser2021}, a negative sign was typographically missed in front of the first term on the right-hand side of Eq.~\eqref{eq:stresseqn}. With the sign displayed in Eq.~\eqref{eq:stresseqn} above, the turbulent stresses are given by
\begin{subequations}
\begin{align}
    \tau_u &= \langle \widetilde{u}_x \widetilde{u}_z\rangle_x = -\langle \partial_z \widetilde{\phi}\cdot \partial_x \widetilde{\phi}\rangle_x  ,\\
    \tau_b &= -\frac{1}{\Ma^2}\langle \widetilde{b}_x \widetilde{b}_z\rangle_x=\frac{1}{\Ma^2}\langle \partial_z \widetilde{\psi} \cdot\partial_x \widetilde{\psi}\rangle_x.
\end{align}
\end{subequations}

These stresses are evaluated from nonlinear simulations and shown in Fig.~\ref{fig:fig4}. Fluctuations of Reynolds stress are concentrated in the shear layer, near $z \approx 0$. Time histories of the Reynolds and Maxwell stresses, at $z=0$, where they are largest in magnitude, are compared in Figs.~\ref{fig:fig4}(b) and \ref{fig:fig4}(d). Note the recurring dominant up-gradient transport via the Reynolds stress. The Maxwell stress, however, is almost always down-gradient. Figures~\ref{fig:fig4}(a) and \ref{fig:fig4}(c) also convey that the Maxwell stress is generally broader along the $z$-axis than the Reynolds stress, which is more localized near the shear layer.

%%%%%%%%%%%%%%%%%%%%%%%%%%%%%%%%%%%%%%%%%%%%%%%%%%%%%%%%%%%%%%%%%%%%%%%%%%%%%%%%%%%%%%%%%%%%%%%%%%%%%
%%%%%%%%%%%%%%%%%%%%%%%%%%%%%%%%%%%%%%%%%%%%%%%%%%%%%%%%%%%%%%%%%%%%%%%%%%%%%%%%%%%%%%%%%%%%%%%%%%%%%
\section{Decomposition of nonlinear simulation onto linear modes}\label{sec:sec5}

To probe the nonlinear simulation data, the turbulent fluctuations are expanded on the linear eigenmode basis described in Sec.~\ref{sec:sec3}. Consider an arbitrary turbulent fluctuation 
$\widetilde{\chi}_{\mathrm{turb}}=
(\widetilde{\phi}_{\mathrm{turb}}, \widetilde{\psi}_{\mathrm{turb}})$, which is expanded as  \begin{equation}
\widetilde{\chi}_{\mathrm{turb}}(x, z, t) = \sum_{k_x\neq 0} \textrm{e}^{ik_x x}\sum_j \beta_j(k_x, t) \chi_j(k_x, z)
\end{equation}
where the eigenmode basis $\chi_j(k_x, z)$ is employed along the $z$-axis at each wavenumber $k_x$ to decompose the fluctuations. \BT{The complex mode-amplitude $\beta_j(k_x,t)$, defined for each eigenmode $j$, can then be computed using properties of the linear operator, described in the Appendix~A, even when the eigenmodes of the operator are non-orthogonal, as is the case here.}

Following earlier studies, \cite{makwana2012, fraser2017, fraser2018, fraser2021, tripathi2022} $j=1,2$ will be used to represent unstable and stable modes, respectively. The computations herein resolve as many as $3109$ eigenmodes at a particular $k_x$.

\subsection{Nonlinear excitation of stable modes}
\begin{figure*}[htbp!]
	\includegraphics[width=1\textwidth]{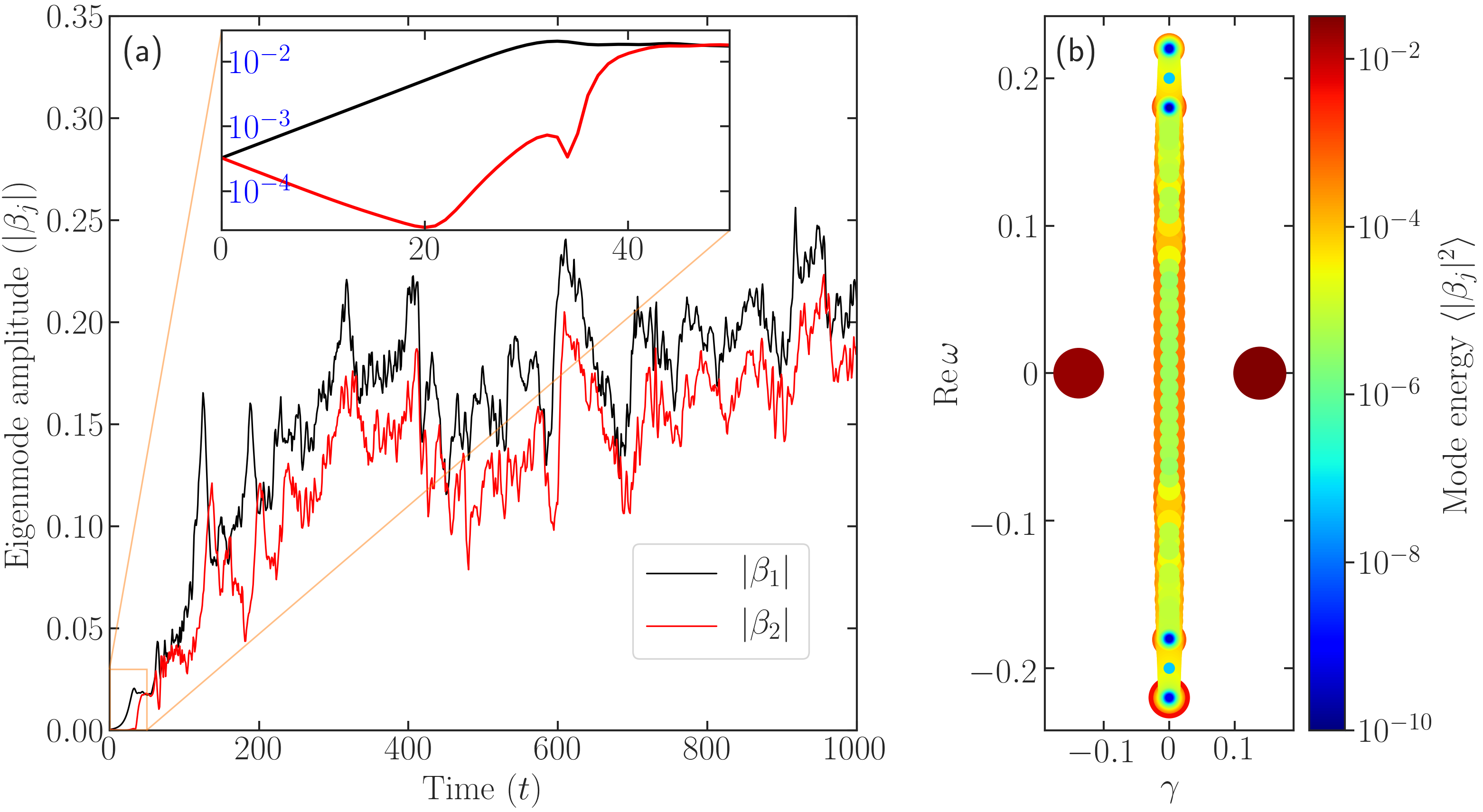}
	\caption{(a) Time traces of the eigenmode amplitudes are shown for $k_x=0.2$ for a simulation with $M_\mathrm{A}=10$ and $D_\mathrm{Krook}=2$. Note, in the inset, the nonlinear excitation of linearly stable mode ($|\beta_2|$) in instability saturation ($|\beta_1|$ is the unstable mode amplitude). (b) All 3109 eigenmodes at $k_x=0.2$ are plotted with their squared excitation levels in the nonlinearly saturated phase, which represent the energy in each eigenmode. The diameter of each circle shown corresponds to the energy in each eigenmode, and modes with lower energy are plotted on top of more highly excited modes, such that all data points are (partially) visible. Note that the total fluctuation energy is composed of both the modal and non-modal energy because of the non-orthogonal modes. Evaluating total energy at a wavenumber, $E = \int dz \left[|\mathbf{u}|^2 + |\mathbf{B}|^2/\Ma^2 \right]/2 = \int dz \left[ (\sum_m \beta_m \mathbf{u}_m)\cdot (\sum_n \beta_n \mathbf{u}_n)^\ast + (\sum_m \beta_m \mathbf{B}_m) \cdot (\sum_n \beta_n \mathbf{B}_n)^\ast /\Ma^2 \right]/2= \sum_{m,n} E_{mn}$, where $(\mathbf{u}_m, \mathbf{B}_m)$ represents the $m$-th eigenmode. When $m$ and $n$ belong to discrete (d) modes, $E_\mathrm{dd}$ is, upon time-averaging ($t=150\textrm{--}1000$), around $72\%$ of the total energy, whereas when $m$ and $n$ belong to continuum (c) modes, $E_\mathrm{cc}$ is $\approx 22\%$; $E_\mathrm{dc}$ is $\approx 6\%$. }\label{fig:fig5}
\end{figure*}

The amplitudes of the unstable and stable modes are tracked in the nonlinear simulations, and their time series are plotted in Fig.~\ref{fig:fig5}(a). As expected, the unstable mode grows and the stable mode decays exponentially in the early phase. However, as the fluctuations increase due to the growth of the unstable modes, nonlinear interactions among them begin exciting the linearly stable mode,\cite{terry2006} causing it to rise to almost the same level as the unstable mode at that wavenumber, see Fig.~\ref{fig:fig5}(a). Later, in the fully nonlinear stage, all eigenmodes can participate in the energy redistribution.

The energy in individual eigenmodes $|\beta_j|^2$, averaged over a turbulent state ($t=150\textrm{--}1000$), is displayed in Fig.~\ref{fig:fig5}(b). It is evident that the unstable and stable eigenmode pair contains a majority ($>70\%$) of the energy in the system. The remaining eigenmodes share a wide spectrum of the remaining energy. This suggests that the turbulent system at hand may be amenable to a substantial dimensionality reduction.\cite{fraser2018, pueschel2016} For the cases of the weaker magnetic fields, this finding is more prominent, as evidenced in the Appendix~B. In addition, the weaker fields support more coherent amplitude-oscillations, unlike the large excursions in the amplitudes observed with the stronger fields, e.g., $\Ma=10$ in Fig.~\ref{fig:fig5}(a). In the latter case, the stronger Lorentz back-reaction acting on the large-scale turbulent flow cause strong oscillations in the eigenmode amplitudes.

%%%%%%%%%%%%%%%%%%%%%%%%%%%%%%%%%%%%%%%%%%%%%%%%%%%%%%%%%%%%%%%%%%%%%%%%%%%%%%%%%%%%%%%%%%%%%%%%%%%%%
%%%%%%%%%%%%%%%%%%%%%%%%%%%%%%%%%%%%%%%%%%%%%%%%%%%%%%%%%%%%%%%%%%%%%%%%%%%%%%%%%%%%%%%%%%%%%%%%%%%%%
% {\color{red}
% \subsection{Why does the vortex nutate?}

% \begin{figure*}[htbp!]
% 	\includegraphics[width=1\textwidth]{PoP/figures/fig6_paper2_600dpi_v1.png}
% 	\caption{MA=60, betas at kx=0.2, vortex nutation}\label{fig:fig6}
% \end{figure*}
% }
	
% \begin{figure*}[htbp!]
% 	\includegraphics[width=1\textwidth]{compare__allkx_phi_at_t_eq_850.png}
% 	\caption{}\label{fig:reconstructionphi}
% \end{figure*}

% \begin{figure}[htbp!]
% 	\includegraphics[width=0.5\textwidth]{relative_error_in_all_kx_phi_upto_t_eq_1000.eps}
% 	\caption{}\label{fig:errorphi}
% \end{figure}

%%%%%%%%%%%%%%%%%%%%%%%%%%%%%%%%%%%%%%%%%%%%%%%%%%%%%%%%%%%%%%%%%%%%%%%%%%%%%%%%%%%%%%%%%%%%%%%%%%%%%
%%%%%%%%%%%%%%%%%%%%%%%%%%%%%%%%%%%%%%%%%%%%%%%%%%%%%%%%%%%%%%%%%%%%%%%%%%%%%%%%%%%%%%%%%%%%%%%%%%%%%s
\subsection{Reduced representation of the turbulent flow} \label{sec:redflow}

To obtain a better understanding of turbulent dynamics, it is of interest to compare different components of eigenmodes in the turbulent flow. An approximate (reduced) representation of the turbulent flow $\widetilde{\phi}_{\mathrm{approx}}$ can be constructed from a class of eigenmodes at each wavenumber, e.g., $\widetilde{\phi}_{\mathrm{approx}}(x,z,t)$ can be written as a sum of an unstable mode per wavenumber $ \sum_{k_x} \textrm{e}^{ik_x x}\beta_1(k_x, t) \phi_1(k_x, z) $, or as a sum of an unstable and a stable mode per wavenumber $ \sum_{k_x} \textrm{e}^{ik_x x}\left[\beta_1(k_x, t) \phi_1(k_x, z)+\beta_2(k_x, t) \phi_2(k_x, z) \right] $. Respective short-hand notations $\beta_1 \phi_1 $ and $\beta_1 \phi_1 +\beta_2 \phi_2 $ will be used hereafter, i.e.,
\begin{subequations}
\begin{align}
    \beta_1 \phi_1 &\equiv \sum_{0<|k_x|<1} \textrm{e}^{ik_x x}\beta_1(k_x, t) \phi_1(k_x, z),\\
    \beta_1 \phi_1 +\beta_2 \phi_2 &\equiv\sum_{0<|k_x|<1} \textrm{e}^{ik_x x}\left[\beta_1(k_x, t) \phi_1(k_x, z)+\beta_2(k_x, t) \phi_2(k_x, z) \right].
\end{align}
\end{subequations}

\begin{figure*}[htbp!]
	\includegraphics[width=1\textwidth]{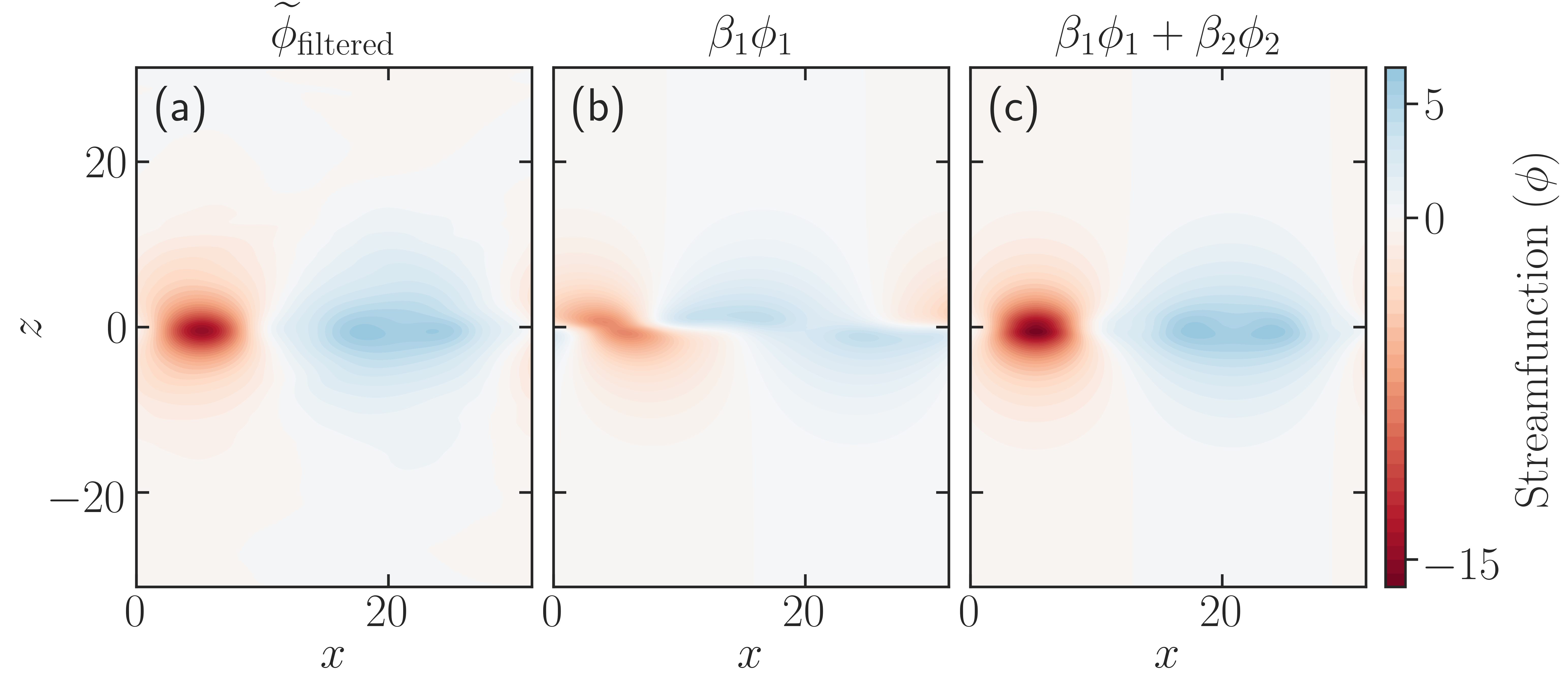}
	\caption{(a) Full turbulent fluctuations in streamfunction in the Kelvin-Helmholtz-unstable wavenumbers $0 < |k_x| < 1$ (thus called $\widetilde{\phi}_\mathrm{filtered}$), observed in nonlinear simulations with $M_\mathrm{A}=30$, $D_\mathrm{Krook}=2$. The shown plot of fluctuations includes all types of eigenmodes---the unstable, stable and continuum modes. (b) Reconstruction of the turbulent fluctuations by summing only the unstable modes at the same wavenumber range. (c) Reconstruction by adding stable and unstable modes, while omitting all continuum modes. The reconstruction in (c) is clearly much alike the turbulent fluctuations in (a), in contrast to the reconstruction in (b). Saturation theory of instability that considers the unstable modes only, at best, can produce (b), but with inclusion of the stable modes, substantial improvement can be achieved.}\label{fig:fig6}
\end{figure*}

The nonlinear fluctuations of the flow are compared in Fig.~\ref{fig:fig6}, viewed at different levels of truncation in the eigenmode expansion. The leftmost panel, Fig.~\ref{fig:fig6}(a), shows the full turbulent fluctuations in the Kelvin-Helmholtz (KH-)unstable wavenumbers $k_x=0.2, 0.4, 0.6, 0.8$, which appear similar to the full turbulent fluctuations that include all wavenumbers in the nonlinear simulation (not shown); Fig.~\ref{fig:fig6}(b) displays the sum of unstable eigenmodes at each of these KH-unstable wavenumbers; and Fig.~\ref{fig:fig6}(c) presents the sum of unstable and stable eigenmodes at the same wavenumbers, while omitting all continuum modes. Adding stable modes produces a substantial improvement in the reconstruction. Note that such a reconstruction was found to deteriorate quickly over time \BT{(i.e., a few instability $\mathrm{e}$-folding times where one $\mathrm{e}$-folding time for the fastest growing mode $k_x=0.4$ is $\gamma^{-1} \approx 5$)} in the study of unforced shear layers, \cite{fraser2021} as the rapid relaxation of the layer towards a stable profile rendered the unstable and stable eigenmodes of the system to be less representative of the decaying turbulence. \BT{The turbulent fluctuation shown in 
Fig.~\ref{fig:fig6} is at $t = 702$, which lies well within the nonlinear phase (the linear phase ends around $t \approx 30$).} In this respect, the forced shear layer is markedly different from the freely evolving layer.

%%%%%%%%%%%%%%%%%%%%%%%%%%%%%%%%%%%%%%%%%%%%%%%%%%%%%%%%%%%%%%%%%%%%%%%%%%%%%%%%%%%%%%%%%%%%%%%%%%%%%
%%%%%%%%%%%%%%%%%%%%%%%%%%%%%%%%%%%%%%%%%%%%%%%%%%%%%%%%%%%%%%%%%%%%%%%%%%%%%%%%%%%%%%%%%%%%%%%%%%%%%s
\subsection{Performance of reduced representations}

While the qualitative analysis of the turbulent-flow reconstruction in Sec.~\ref{sec:redflow} is instructive, a quantitative measurement is desirable. To this end, following Ref.~\cite{fraser2021}, the reconstructive capability of reduced representations is quantified, at each time step in the simulation, using the standard energy norm that measures the fraction of kinetic energy lost when the eigenmode basis is truncated, compared to the kinetic energy in the full turbulent flow data---see the definition in Eq.~\eqref{eq:truncationerror}. The energy norm is well-suited for studying large-scale structures. Small-scale phenomena, however, may not be amenable to such analysis, although one may be able to find ties between the small- and large-scale pheonomena in some cases. This measure is also called a ``truncation error." Note that this error arises not in the nonlinear simulations but merely in the reduced representations of turbulent fluctuations, when truncating the eigenmode basis in post-processing analyses.
\begin{figure*}[htbp!]
	\includegraphics[width=1\textwidth]{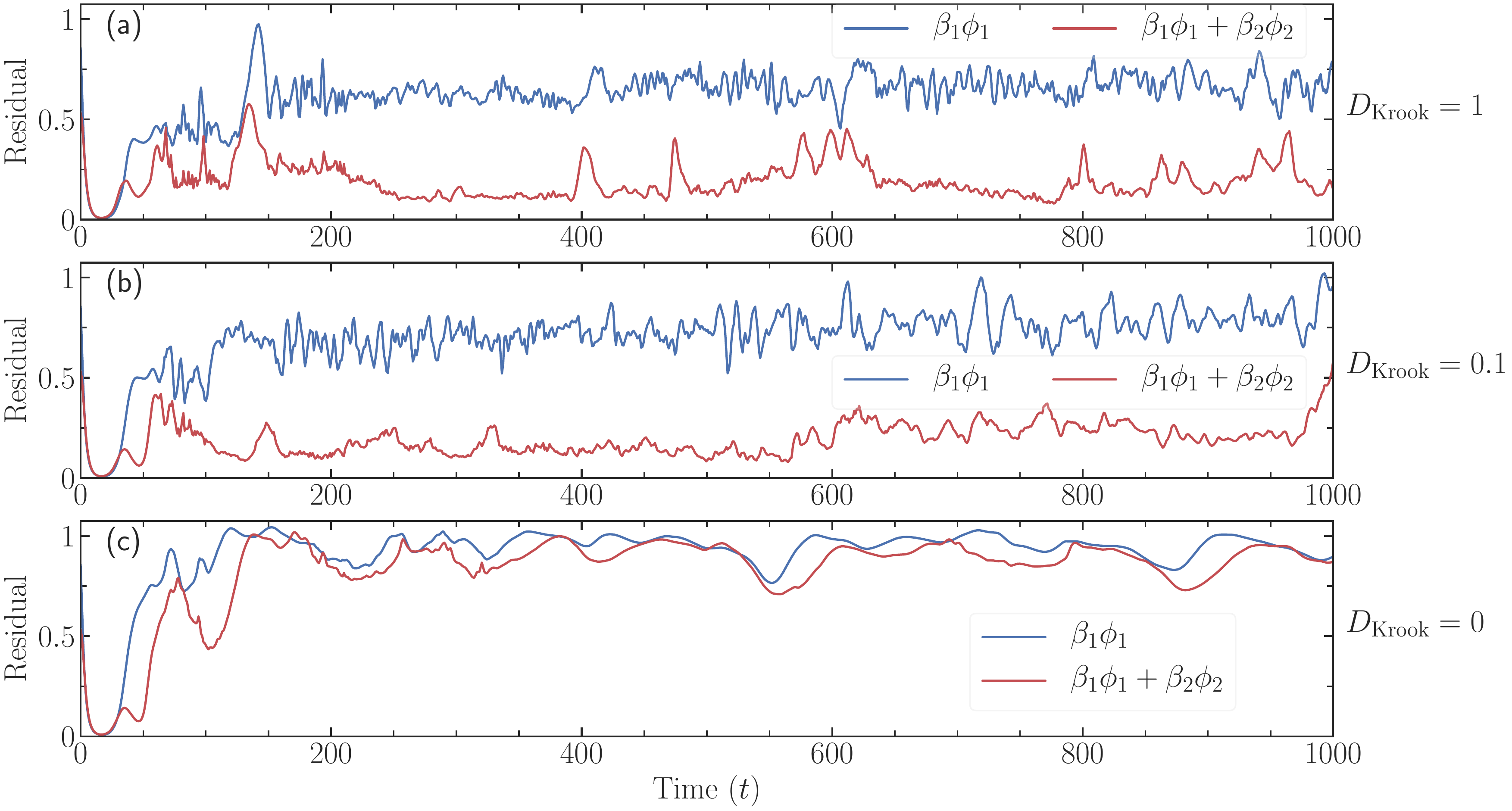}
	\caption{Time traces of residuals, i.e., the fraction of energy missed in truncated bases, normalized to the total energy in the turbulent flow at each time step. The reconstruction uses truncated bases with unstable modes alone, and unstable and stable modes together. The forcing strength is varied in three different simulations with $\Ma=10$: (a) $D_\mathrm{Krook}=1$, (b) $D_\mathrm{Krook}=0.1$, and (c) $D_\mathrm{Krook}=0$. The unforced shear layer in (c) rapidly flattens out, and thus instability no longer drives the turbulence. As long as the turbulence is driven by the instability, the unstable and stable modes together can reconstruct a large fraction of the turbulent flow features in (a) and (b). }\label{fig:fig7}
\end{figure*}

Using the energy norm, we define the relative truncation error, which may also be called a normalized residual, in the following manner:
\begin{equation} \label{eq:truncationerror}
\begin{aligned}
    \mathrm{Residual} = \frac{||\widetilde{\phi}_{\mathrm{exact}} - \widetilde{\phi}_{\mathrm{approx}}||_{\mathrm{KE}}^2}{||\widetilde{\phi}_{\mathrm{exact}}||_{\mathrm{KE}}^2} = \frac{||\widetilde{\phi}_\mathrm{diff}||_{\mathrm{KE}}^2}{||\widetilde{\phi}_{\mathrm{exact}}||_{\mathrm{KE}}^2}=\frac{\int dxdz \left[\left(\partial_x \widetilde{\phi}_{\mathrm{diff}}\right)^2 + \left(\partial_z \widetilde{\phi}_{\mathrm{diff}}\right)^2 \right]  }{\int dxdz \left[\left(\partial_x \widetilde{\phi}_{\mathrm{exact}}\right)^2 + \left(\partial_z \widetilde{\phi}_{\mathrm{exact}}\right)^2 \right] },
\end{aligned}
\end{equation}
where $(\partial_x \phi)^2$ and $(\partial_z \phi)^2$ are the squared $z$- and $x$-components of velocities; $\widetilde{\phi}_\mathrm{diff} = \widetilde{\phi}_{\mathrm{exact}} - \widetilde{\phi}_{\mathrm{approx}}$ with $\widetilde{\phi}_{\mathrm{exact}} $ and $\widetilde{\phi}_{\mathrm{approx}} $ representing respectively the turbulent streamfunction from nonlinear simulation and its reduced representation---either a summation over the unstable modes alone or over the unstable and stable modes together---both spanning fluctuations over a range of wavenumbers. Here, this range, taken to be the same for both, is considered to be $0 < |k_x|  < 1$, which corresponds to the wavenumber range of the instability. If the residual is less than unity, the truncation in the eigenmode expansion may be considered as a representative of the full system and thus a candidate for reduced-order model building. On the contrary, the residual being around unity or more signfies the failure of the reduced representation in effectively capturing the overall nonlinear fluctuations.
\begin{figure}[htbp!]
	\includegraphics[width=0.48\textwidth]{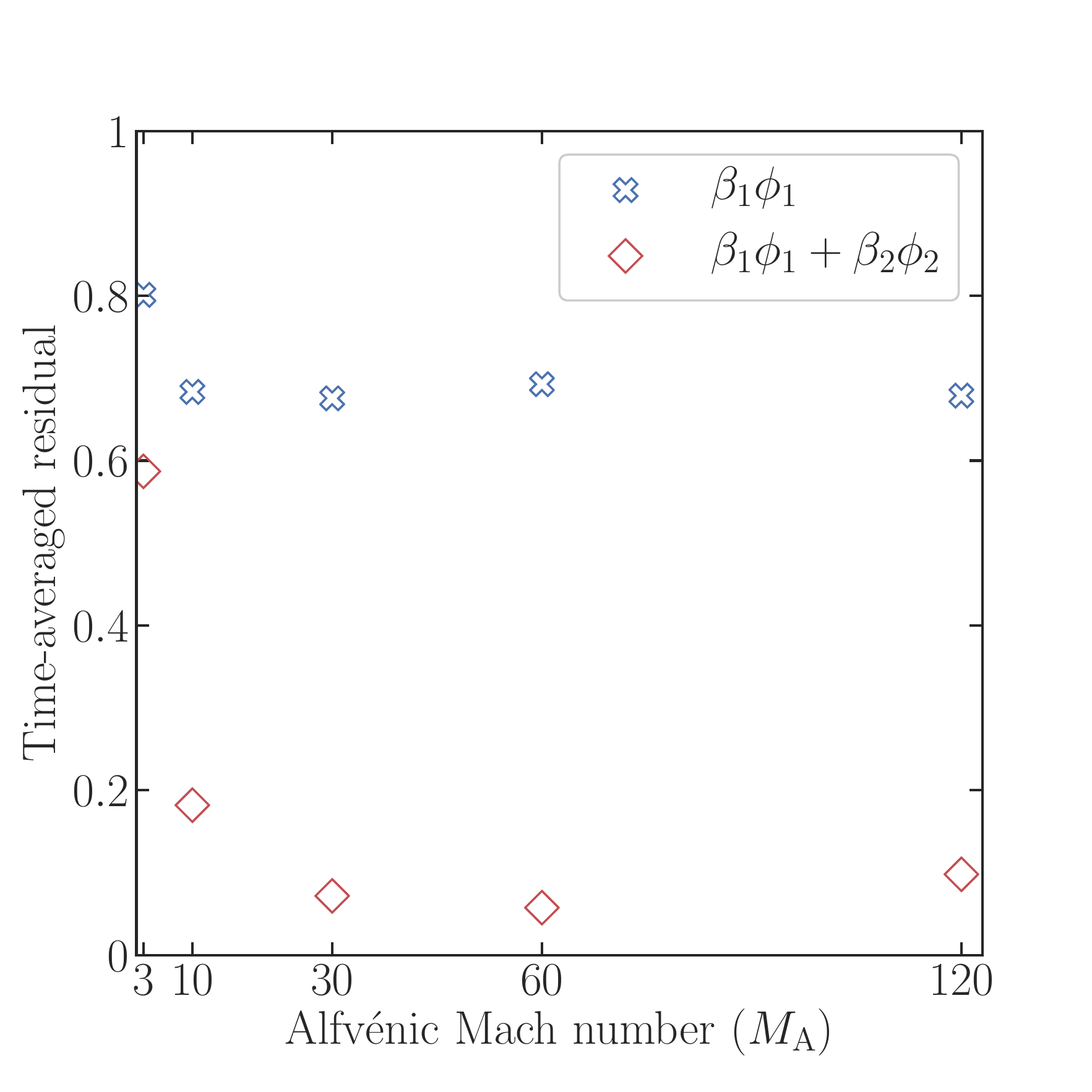}
	\caption{Shown are the time-averaged residuals for simulations different $\Ma$. Note the residuals are the fractions of energy missed in the truncated bases, compared to the total energy in the instantaneous full turbulent flow. Time-averaging is performed over a quasi-stationary state of turbulence $(t=600\text{--}1000)$. The reconstruction uses truncated bases with unstable modes alone, and unstable and stable modes together, leaving all the continuum modes. All simulations use $D_\mathrm{Krook}=2$. Note the dramatic improvement with the inclusion of the stable modes. For $\Ma=3$, the improvement is modest, as the stronger Lorentz force back-reacts on the large-scale turbulent flow, producing more fluctuations in the continuum modes.}\label{fig:fig8}
\end{figure}

The time evolution of the residuals is compared in Fig.~\ref{fig:fig7} for varying forcing strengths. As expected, the unstable modes entirely capture the fluctuations in the linear phase (i.e., $t \lesssim 30$). In the nonlinear phase, however, the unstable modes capture only a rather limited fraction of the turbulent fluctuations. This is greatly improved when the stable modes are added. This suggests that the success of quasilinear models in capturing key properties of the turbulence can crucially depend on whether stable modes are considered when constructing such models.

It is also interesting to note that the turbulence in the unforced shear layer, see Fig.~\ref{fig:fig7}(c), is different from the forced cases. In the former, the shear layer quickly flattens out and nearly shuts off the instability, leading to a decaying turbulence. Regardless of whether the unstable and/or stable modes are considered, the corresponding reconstructions fail to model the turbulence with any degree of accuracy. By contrast, when the shear layer is forced, a reduced representation of the turbulent flow with two modes (per wavenumber) is found to perform well, recovering a substantial fraction of the full nonlinear system.

A similar reconstruction is shown for various strengths of magnetic fields $\Ma$ in Fig.~\ref{fig:fig8}, where residuals are time-averaged over a quasi-stationary state of turbulence. With stronger magnetic fields (lower $M_\mathrm{A}$), the vortices begin disrupting due to stronger Lorentz force and consequently generate more fluctuations at scales beyond the Kelvin-Helmholtz-instability (KHI) range. \cite{mak2017} This accounts for an increase of the residual for low $M_\mathrm{A}$, although it remains below $0.2$ for $M_\mathrm{A}=10$. For $\Ma=3$, the improvement with the inclusion of the stable modes is modest. Momentum transport by large-scale structures, formed from the unstable and stable modes, within the KHI range, however, may still dominate over the transport contributed by much smaller scales or the remaining continuum modes; hence, a quantitative analysis of transport will be conducted next.
%%%%%%%%%%%%%%%%%%%%%%%%%%%%%%%%%%%%%%%%%%%%%%%%%%%%%%%%%%%%%%%%%%%%%%%%%%%%%%%%%%%%%%%%%%%%%%%%%%%%%
%%%%%%%%%%%%%%%%%%%%%%%%%%%%%%%%%%%%%%%%%%%%%%%%%%%%%%%%%%%%%%%%%%%%%%%%%%%%%%%%%%%%%%%%%%%%%%%%%%%%%s
\subsection{Competing up- and down-gradient momentum transport and their reduced models}\label{sec:transportmodeling}

The Reynolds stress can be expressed in terms of the contribution from each wavenumber, which can further be decomposed into the contribution from each eigenmode. At a wavenumber $k_x$, the Reynolds stress from all the fluctuations $\hat{\phi}_{k_x}$ read 
\begin{equation} \label{eq:turbdnsu}
    \tau_u(\mathrm{all\ modes}) = 2 \mathrm{\ Im} [k_x \hat{\phi}_{k_x} \cdot \partial_z\hat{\phi}_{k_x}^{\ast} ],
\end{equation}
whereas the contribution from an unstable mode alone, and from an unstable and a stable mode alone, at that wavenumber are respectively given as  
\begin{equation}\label{eq:unstabletrans}
\begin{aligned}[b]
    \tau_u(\mathrm{unstable}) &= 2\mathrm{\ Im} [k_x (\beta_1 \phi_{1,k_x}) \cdot \partial_z (\beta_1 {\phi}_{1,k_x})^{\ast} ]\\
    &= 2|\beta_1|^2 \mathrm{\ Im} [k_x \phi_{1,k_x} \cdot \partial_z {\phi}_{1,k_x}^{\ast} ],
\end{aligned}
\end{equation}
and
\begin{equation} \label{eq:stabletrans}
\begin{aligned}[b]
    \tau_u(\mathrm{stable}) &= 2|\beta_2|^2 \mathrm{\ Im} [k_x \phi_{2,k_x} \cdot \partial_z{\phi}_{2,k_x}^{\ast} ]\\
    &= 2|\beta_2|^2 \mathrm{\ Im} [k_x \phi_{1,k_x}^\ast \cdot \partial_z{\phi}_{1,k_x} ]\\
    &= -2|\beta_2|^2 \mathrm{\ Im} [k_x \phi_{1,k_x} \cdot \partial_z{\phi}_{1,k_x}^\ast ],
\end{aligned}
\end{equation}
where $\hat{\phi}_{k_x}$ is the Fourier transform of the streamfunction at wavenumber $k_x$ and $\phi_{j,k_x}$ represents the $z$-dependent $j$-th complex eigenmode: $j=1,2$ for unstable and stable modes, respectively. The conjugate symmetry of unstable and stable modes, as shown in Figs.~\ref{fig:fig1}(b) and ~\ref{fig:fig1}(c), is used in Eq.~\eqref{eq:stabletrans}, i.e., $\phi_{2,k_x}(z) = \phi_{1,k_x}^\ast(z)$. The negative sign of the last expression in Eq.~\eqref{eq:stabletrans} corresponds to the up-gradient nature of momentum transport by stable modes, which was physically analyzed in Sec.~\ref{sec:competeunstablestable} and visually demonstrated in Fig.~\ref{fig:fig2}.

The summed contributions of unstable and stable modes in transport, however, can have cross-terms---quadratic correlations between unstable and stable modes---that do not appear in Eqs.~\eqref{eq:unstabletrans} and \eqref{eq:stabletrans} where contribution from individual modes are shown. But the cross-terms vanish when the unstable and stable modes are \textit{exactly} complex conjugates of each other, as is the case for the ideal Kelvin-Helmholtz instability \BT{ (when this conjugate symmetry is broken, e.g., in resistive tearing instability or in ion-temperature-gradient instability,\cite{terry2009} the cross-terms can have non-zero contribution)}:
\begin{equation}
\begin{aligned}[b]
    \tau_u(\mathrm{unstable\ + \ stable}) &= 2\mathrm{\ Im} [k_x (\beta_1 \phi_{1,k_x}+\beta_2 \phi_{2,k_x}) \cdot \partial_z (\beta_1 {\phi}_{1,k_x}+\beta_2 \phi_{2,k_x})^{\ast} ]\\
    &=2 \left(|\beta_1|^2-|\beta_2|^2\right) \mathrm{\ Im} [k_x {\phi}_{1,k_x} \cdot \partial_z{\phi}_{1,k_x}^{\ast} ] + \mathrm{cross\textnormal{-}terms},
\end{aligned}
\end{equation}
where
\begin{equation} \label{eq:crosstermsvanish}
\begin{aligned}[b]
   \mathrm{cross\textnormal{-}terms} &= 2\mathrm{\ Im} [k_x (\beta_1 \phi_{1,k_x}) \cdot \partial_z (\beta_2 \phi_{2,k_x})^{\ast} ] + 2\mathrm{\ Im} [k_x (\beta_2 \phi_{2,k_x}) \cdot \partial_z (\beta_1 \phi_{1,k_x})^{\ast} ]\\
    &= 2\mathrm{\ Im} [\beta_1 \beta_2^\ast k_x  \phi_{1,k_x} \cdot \partial_z  \phi_{2,k_x}^{\ast} ] + 2\mathrm{\ Im} [\beta_2 \beta_1^\ast k_x \phi_{2,k_x} \cdot \partial_z  \phi_{1,k_x}^{\ast} ]\\
    &= 2\mathrm{\ Im} [\beta_1 \beta_2^\ast k_x  \phi_{1,k_x} \cdot \partial_z  \phi_{2,k_x}^{\ast} ] + 2\mathrm{\ Im} [\beta_2 \beta_1^\ast k_x \phi_{1,k_x}^\ast \cdot \partial_z  \phi_{2,k_x} ]\\
    &= 2\mathrm{\ Im} [\beta_1 \beta_2^\ast k_x  \phi_{1,k_x} \cdot \partial_z  \phi_{2,k_x}^{\ast} ] + 2\mathrm{\ Im} [\left(\beta_1 \beta_2^\ast k_x  \phi_{1,k_x} \cdot \partial_z  \phi_{2,k_x}^{\ast} \right)^\ast]\\
    &=0.
\end{aligned}
\end{equation}
Thus we obtain
\begin{equation} \label{eq:transsolar}
\begin{aligned}[b]
    \tau_u(\mathrm{unstable\ + \ stable})
    &= 2 \left(|\beta_1|^2-|\beta_2|^2\right) \mathrm{\ Im} [k_x {\phi}_{1,k_x} \cdot \partial_z{\phi}_{1,k_x}^{\ast} ]\\
    &= \tau_u(\mathrm{unstable}) + \tau_u(\mathrm{stable}).
\end{aligned}
\end{equation}

These relations inform us about the $z$-profile of the Reynolds stress, contributed by each wavenumber and each eigenmode. As largest momentum transport happens in the region with the largest flow-gradient, it is instructive to compute, in the forced shear layers, the turbulent stresses at the middle of the layer at $z=0$, and compare the stress contributions from different eigenmodes at various wavenumbers. 

\begin{figure*}[htbp!]
	\includegraphics[width=1\textwidth]{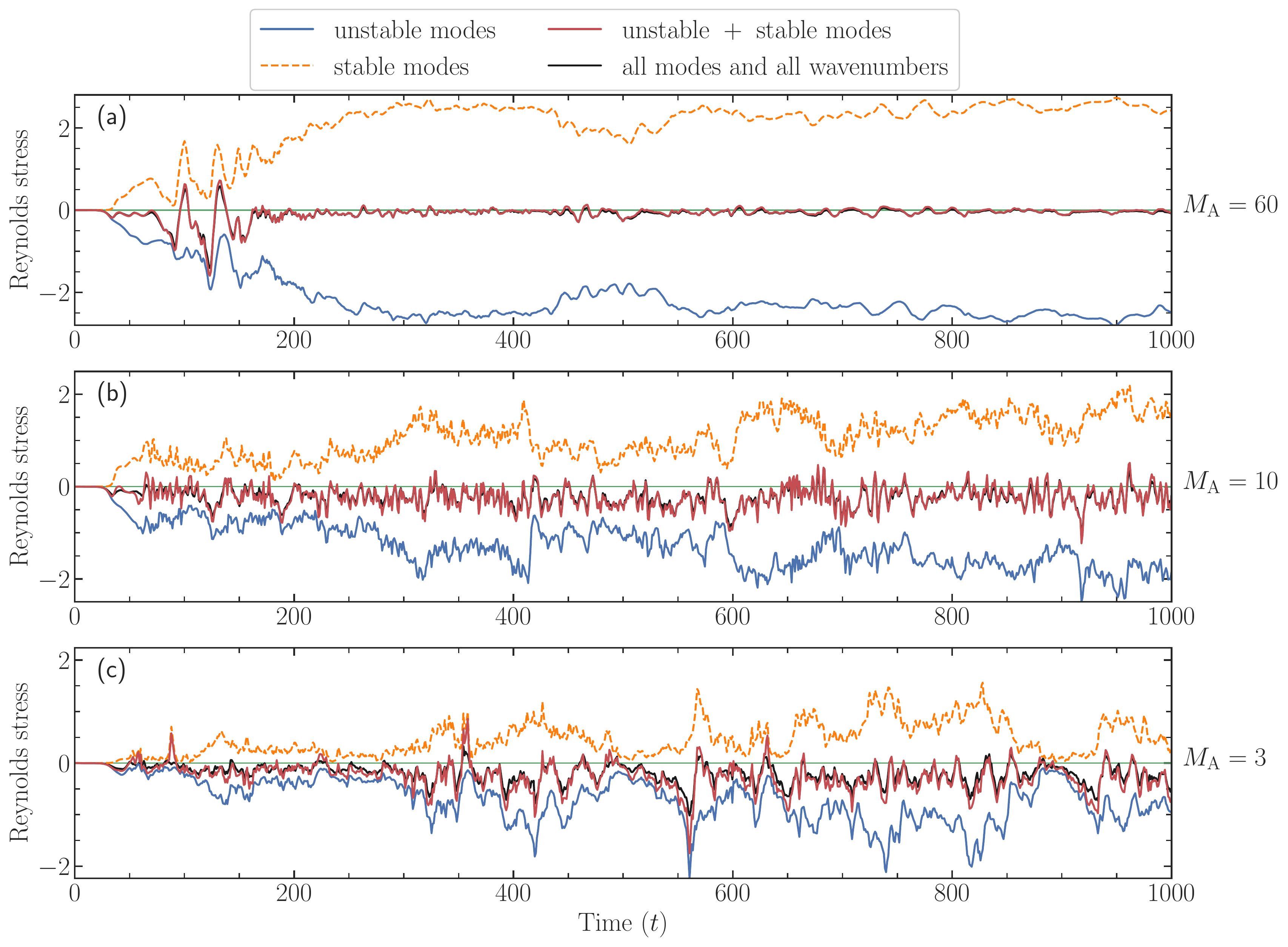}
	\caption{Time variations of Reynolds stress at the middle of the shear layer, $z=0$. The stress contributions from unstable modes (blue), stable modes (orange), their sum (red), and full nonlinear fluctuations, i.e., all modes and all wavenumbers in the simulations (black), are compared, for varying strengths of magnetic fields: (a) $\Ma=60$, (b) $\Ma=10$, and (c) $\Ma=3$. Thin green lines represent the zero level. All simulations use $D_\mathrm{Krook}=2$. Although, with stronger magnetic fields, the up-gradient momentum transport by stable modes are reduced, the up- and down-gradient transport nearly cancel each other throughout all cases.}\label{fig:fig9}
\end{figure*}

The total Reynolds stress from all modes and all wavenumbers in the simulations is compared in Fig.~\ref{fig:fig9} with the stress contributions from the wavenumber range $0 < |k_x| < 1$, which is decomposed further into eigenmodes to assess the contribution of the unstable modes, stable modes, and their sum. The subplots demonstrate that the stable modes are highly efficient in transporting momentum in the up-gradient direction, as compared to the down-gradient transport by the unstable modes. Even for the strongest magnetic field $M_\mathrm{A} = 3$, close to the instability threshold, the stable modes contribute significantly to a \textit{continuous} reduction of the turbulent momentum flux. In addition, the occasional breakthroughs in stable-mode activity cause reversals of the transport direction. This reversal can be observed when the total Reynolds stress in the system is computed, without decomposing the stress into contributions by each eigenmode. However, when the stable modes are not overtaking the unstable modes in transport, the resulting down-gradient transport observed in simulations or experiments is difficult to interpret, in regards to the contributions of stable modes in subdominantly reducing the transport. An eigenmode decomposition of turbulent fluctuations, however, uncovers a complete picture, as is shown here. 
%\BT{This establishes that inclusion of the stable modes in quasilinear models is imperative if one wishes to build a transport model that is reliable. Ignoring stable modes can lead to predictions that are dramatically different from the true transport rates.}
%It should be noted that, often times, in studies of shear-flow turbulence and transport, one does not even distinguish the contributions from unstable modes and remaining fluctuations. Thus the current analysis of unstable and stable modes competing to transport momentum in opposing direction can be insightful to interpret previous experimental and numerical studies, e.g., reduction of transport when a coherent vortex emerges from the shear-flow instability and self-oscillations of the mean flow.}

% Note that the early phases are different from the latter phases of time evolution for weak and strong magnetic fields. In the early phase, stronger fields reduce transport rate. But in the latter phase, when the fields are weak, a self-organization takes over, significantly exciting the stable modes. This understanding is essentially tied to the stable modes that almost completely cancel the unstable modes' transport efficiency in weak field cases and moderately in strong field cases.

\begin{figure*}[htbp!]
	\includegraphics[width=1\textwidth]{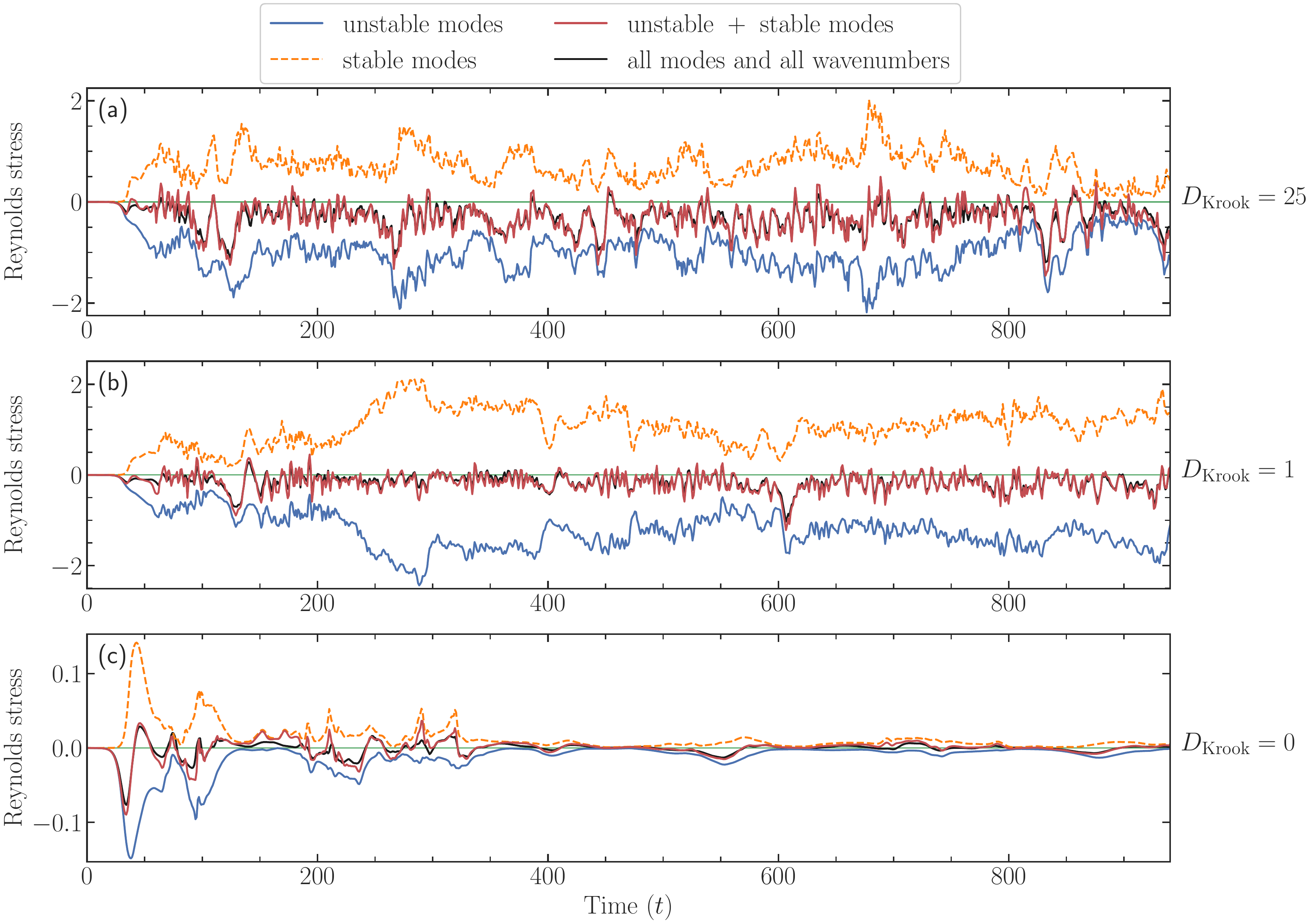}
	\caption{Time variations of Reynolds stress at the middle of the shear layer, $z=0$. The stress contributions from unstable modes (blue), stable modes (orange), their sum (red), and full nonlinear fluctuations, i.e., all modes and all wavenumbers in the simulations (black), are compared, for varying forcing strengths: (a) $D_\mathrm{Krook}=25$, (b) $D_\mathrm{Krook}=1$, and (c) $D_\mathrm{Krook}=0$. All simulations use $\Ma=10$. Thin green lines represent the zero level. Qualitative differences can be observed in unforced ($D_\mathrm{Krook}=0$) and forced cases ($D_\mathrm{Krook}\neq 0$): as instability extracts energy from the mean flow, the profile relaxation in the unforced layer leads to a decaying turbulence, and the transport rates become very small [note the vertical axis labels in (c)]. However, in all cases, the summed stable modes producing up-gradient transport nearly cancel the down-gradient transport by unstable modes. The  addition of these two contributions produces a stress that is almost identical to the stress from all modes. }\label{fig:fig10}
\end{figure*}

Similar variations of momentum transport across the middle of the shear layer are compared in Fig.~\ref{fig:fig10} for different forcing strengths. Note the unforced case differs from the forced cases, as the nearly-flattened shear layer has less momentum to be transported across the layer. As reported in Ref.~\cite{fraser2021}, despite the profile relaxation, the two eigenmodes per wavenumber describe well the temporal variation of the Reynolds stress across the shear layer, although the stress itself is very low (note its vertical scale). In all cases, the stress captured via the sum of unstable and stable modes almost completely follows the total stress from all modes.

\begin{figure*}[htbp!]
	\includegraphics[width=1\textwidth]{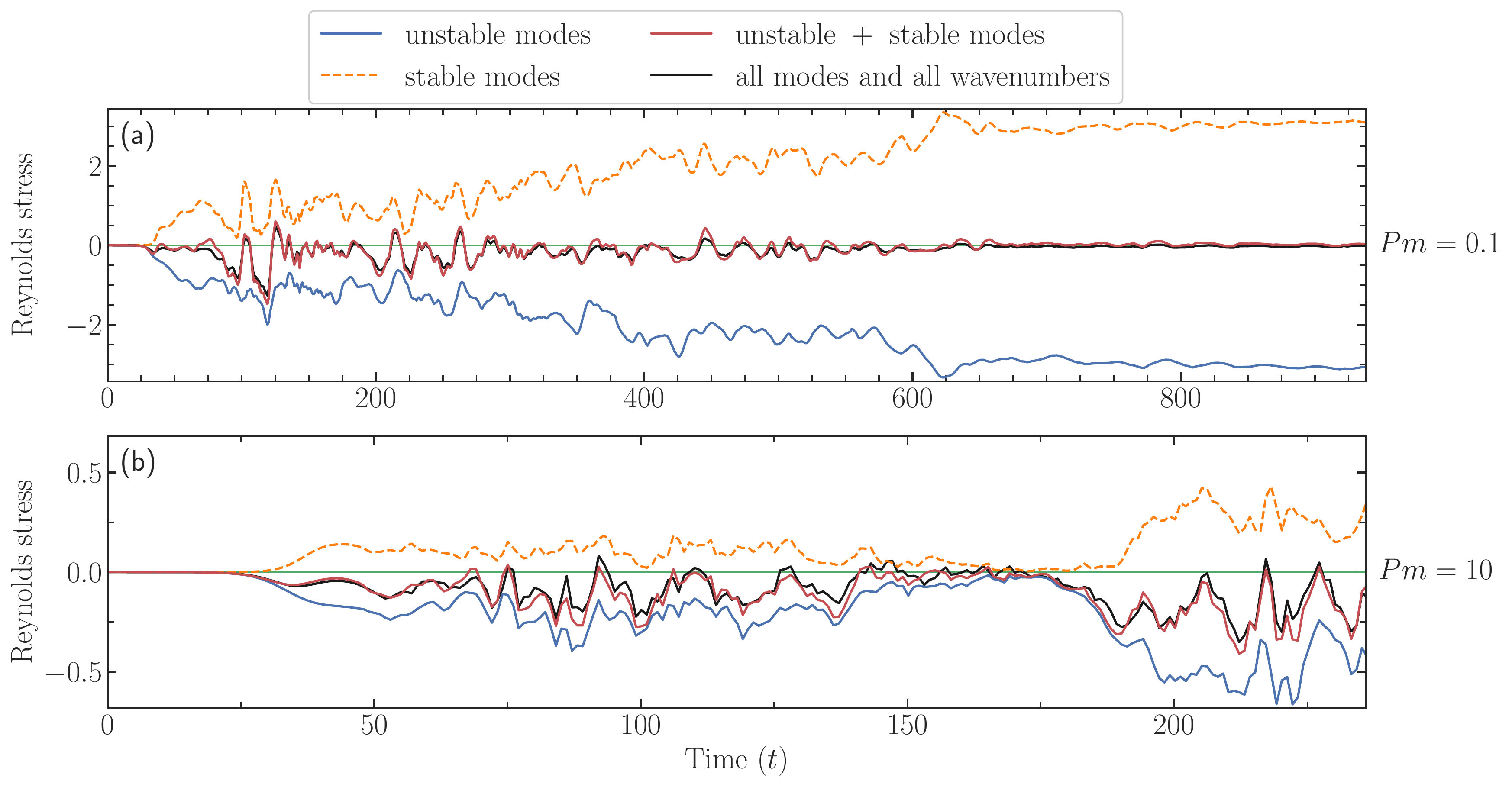}
	\caption{Time variations of Reynolds stress at the middle of the shear layer, $z=0$. The stress contributions from unstable modes (blue), stable modes (orange), their sum (red), and full nonlinear fluctuations, i.e., all modes and all wavenumbers in the simulations (black), are compared, for varying magnetic Prandtl numbers (resistivities): (a) $\Pm=0.1$ and (b) $\Pm=10$. All simulations use $\Ma=10$, $D_\mathrm{Krook}=2$, and $\Re=500$. Thin green lines represent the zero level. It can be observed that the stable modes begin driving up-gradient momentum transport at around $t\approx 30$ when the nonlinear phase of evolution begins. By varying $\Pm$ by two orders of magnitude, around \textit{unity}, the stable modes are found to substantially reduce the down-gradient transport; note the case of $\Pm=1$ is shown in Fig.~\ref{fig:fig9}(b). }\label{fig:fig11}
\end{figure*}

In Fig.~\ref{fig:fig11}, the momentum transport by the unstable and stable modes is presented as a function of magnetic Prandtl number $\Pm=\Rm/\Re$. All simulations until this point used $\Rm=500$, which is now changed to $\Rm=50$ and $\Rm=5\,000$. In both cases of $\Pm=0.1$ and $\Pm=10$, the stable modes still substantially offset the turbulent momentum transport of the unstable modes. The shorter time trace for $\Pm=10$ is due to the higher simulation cost. It should be noted that the quasi-stationary state in this simulation is still undergoing changes, unlike in the case of $\Pm=0.1$ in Fig.~\ref{fig:fig11}(a) or $\Pm=1$ in Fig.~\ref{fig:fig9}(b), all with the same $M_\mathrm{A}=10$, $D_\mathrm{Krook}=2$ and $\Re=500$.

\begin{figure*}[htbp!]
	\includegraphics[width=1\textwidth]{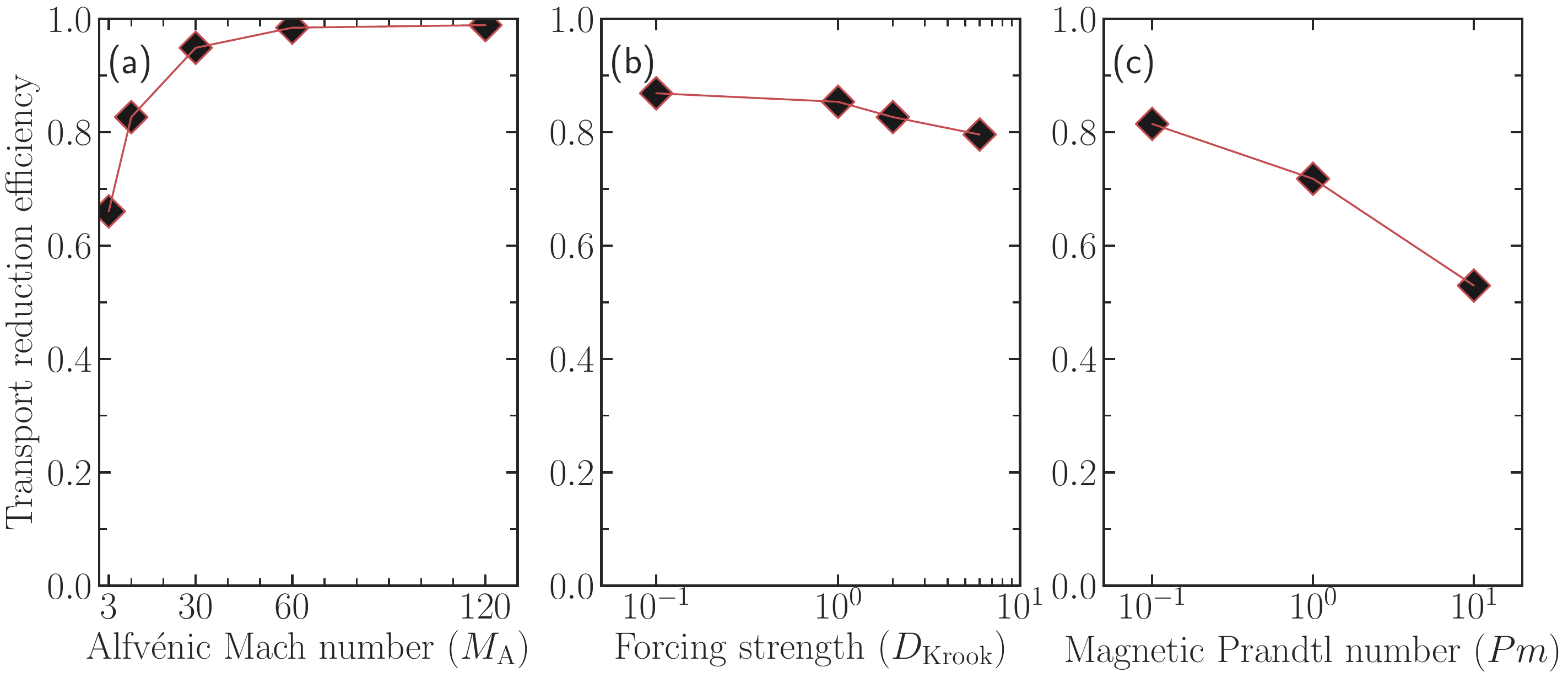}
	\caption{Parameter dependence of transport reduction efficiency, which is defined as the ratio of time-averaged up-gradient Reynolds stress due to stable modes and time-averaged down-gradient Reynolds stress due to unstable modes. The stress is measured at the middle of the shear layer, $z=0$, where the momentum transport is at its maximum. (a) Variations in $M_\mathrm{A}=3, 10, 30, 60, 120$ with $D_\mathrm{Krook}=2$, $\Pm=1$, and linear $x$-scale. (b) Variations in $D_\mathrm{Krook}=0.1, 1,2,6$ with $M_\mathrm{A}=10$, $\Pm=1$, and logarithmic $x$-scale. (c) Variations in $\mathrm{Pm}=0.1,1,10$ (or, equivalent changes in resistivities) with $M_\mathrm{A}=10$, $D_\mathrm{Krook}=2$, and logarithmic $x$-scale. All plots have the same $y$-axis. The time-average for (a) and (b) is taken over a long quasi-stationary state of turbulence $t=350\text{--}900$, while for (c), it is $t=137\text{--}237$ where the quasi-stationary state is still undergoing changes. In all cases, substantial reduction of transport by stable modes is evident, which cancel, via their up-gradient transport, more than half of the down-gradient transport by unstable modes, and this fraction reaches up to $98\%$, see (a), for $\Ma=60$ and $\Ma=120$.} \label{fig:fig12}
\end{figure*}

The efficiency of time-averaged up-gradient momentum transport due to stable modes is compared in Fig.~\ref{fig:fig12} with the time-averaged down-gradient transport due to unstable modes, via a measure, defined below:
\begin{equation}
    \mathrm{Transport\ reduction\ efficiency} = \frac{\langle \mathrm{Up\text{-}gradient\ transport\ by\ stable\ modes}\rangle_t}{\langle\mathrm{Down\text{-}gradient\ transport\ by\ unstable\ modes}\rangle_t},
\end{equation}
where $\langle A \rangle_t$ represents a time-averaging operation on $A$.

Variations in magnetic field strength, forcing strength, and magnetic Prandtl number all demonstrate  that the stable modes cancel an appreciable amount of the turbulent momentum flux associated with the unstable modes. On average, around $80\%$ of the down-gradient flux is offset in this manner.

A remark should be made now regarding the use of unstable and stable modes for building a reliable reduced mode of transport for geo- and astro-physical problems. One approach would be to relate the activity of these two modes with a coefficient of diffusive flux (although the unstable and stable modes offer spatial profiles of transport as well, with both diffusive and non-diffusive fluxes, because they do not rely on an ad-hoc eddy-viscosity model, which is an explicit diffusive-flux-based model). In the middle of the shear layer, the diffusive flux, however, dominates because of the maximum in the flow-gradient. The ad-hoc turbulent viscosity can thus be defined \cite{fraser2018}, more importantly without a ``free-parameter," using Eq.~\eqref{eq:transsolar} as 
\begin{equation} \label{eq:turbvisc}
    \nu_{\mathrm{turb}}=\frac{\tau_u(\mathrm{unstable\ + \ stable})}{(dU_0/dz)\Big\vert_{z=0}}
    = \sum\limits_{0 <k_x < 1} 2 \left(|\beta_1|^2-|\beta_2|^2\right) \mathrm{\ Im} [k_x {\phi}_{1,k_x} \cdot \partial_z{\phi}_{1,k_x}^{\ast} ]\Big\vert_{z=0}.
\end{equation}
Note that the denominator is unity for the shear-flow that has a linear profile in the vicinity of $z=0$. To assess the importance of stable modes in this construct, $|\beta_1|^2-|\beta_2|^2$ can be written as $|\beta_1|^2\left(1-|\beta_2|^2/|\beta_1|^2\right)$. Since $|\beta_2|^2$ has been found to on the same order of $|\beta_1|^2$, e.g., see Fig.~\ref{fig:fig12}, where $|\beta_2|^2/|\beta_1|^2$ can range from $\approx 0.8$ to $\approx 0.95$, yielding $\left(1-|\beta_2|^2/|\beta_1|^2\right) \approx 0.05 \mathrm{\ to\ } 0.2$. Therefore, neglecting stable modes can overestimate the transport by a factor of $5$ to $20$.

\subsection{Reynolds vs. Maxwell stresses}

With the above successful low-order representation of Reynolds stress above, we now examine the fluctuations in the magnetic field that give rise of Maxwell stress. The stress can be quantified as
\begin{equation}
    \tau_b(\mathrm{all\ modes}) = -\frac{2}{M_\mathrm{A}^2} \mathrm{\ Im} [k_x \hat{\psi}_{k_x} \cdot \partial_z\hat{\psi}_{k_x}^{\ast} ],
\end{equation}

\begin{equation}
    \tau_b(\mathrm{unstable\ + \ stable}) =  -\frac{2}{M_\mathrm{A}^2} \left(|\beta_1|^2-\beta_2|^2\right) \mathrm{\ Im} [k_x {\psi}_{1,k_x} \cdot \partial_z{\psi}_{1,k_x}^{\ast} ],
\end{equation}
where $\hat{\psi}_{k_x}$ is the Fourier transform of the flux function at a wavenumber $k_x$ and $\psi_{j,k_x}$ represents the $z$-dependent $j$-th complex eigenmode ($j=1,2$ for unstable and stable modes, respectively). Again, cross-terms arising from the correlation between the unstable and modes can be shown to vanish, exactly as it was shown for the Reynolds stress in Eq.~\eqref{eq:crosstermsvanish}.

\begin{figure*}[htbp!]
	\includegraphics[width=1\textwidth]{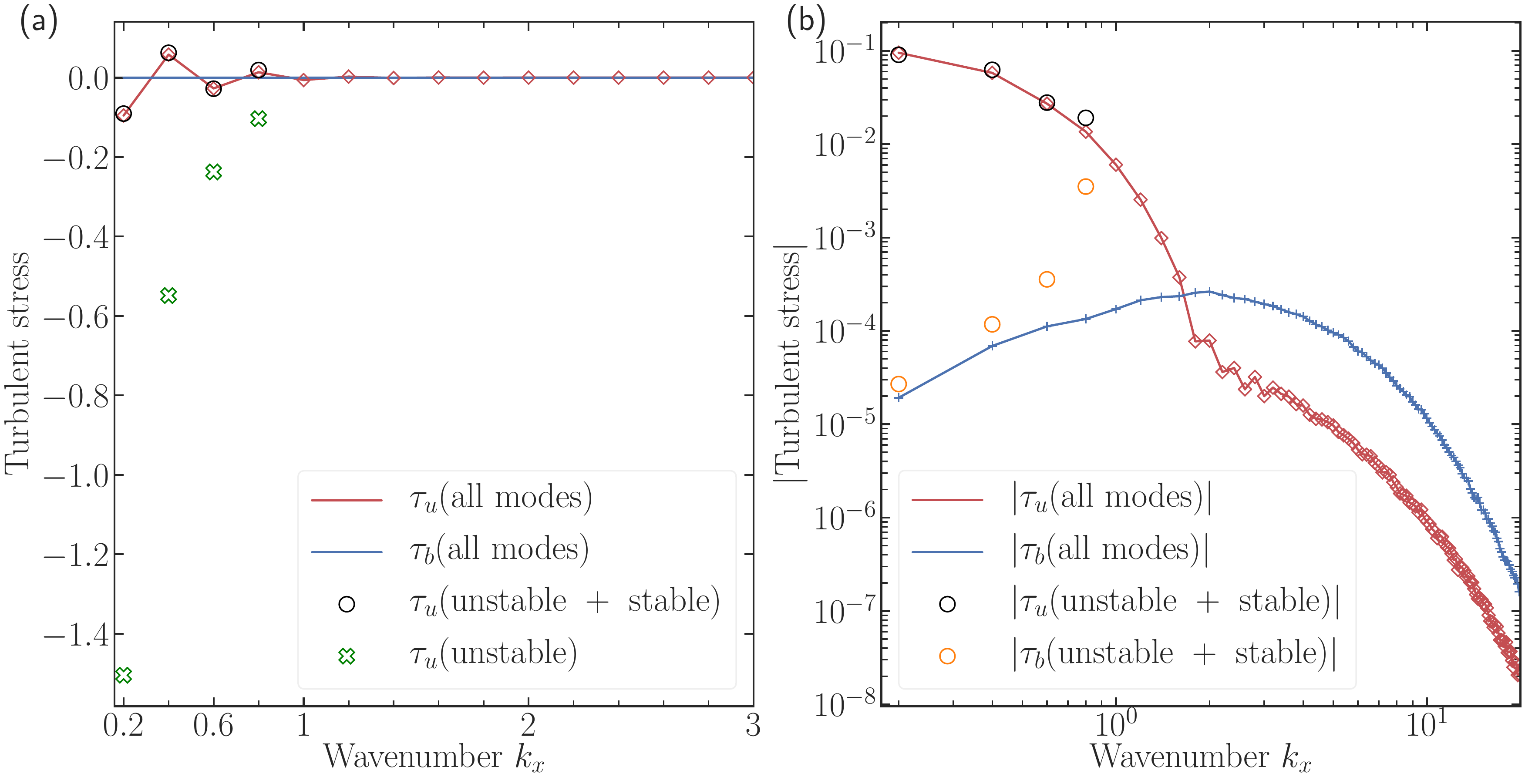}
	\caption{Time-averaged turbulent stresses split into their contributions from different wavenumbers. (a) Stresses on a linear scale. (b) Log-log representation of the absolute value of the stresses. Note that only the wavenumbers $k_x < 1$ are Kelvin-Helmholtz-unstable. The simulation parameters are $M_\mathrm{A}=60$ and $D_\mathrm{Krook}=2$; the time-average is taken over a quasi-stationary state of turbulence, $t=350\text{--}1000$. The total turbulent stress is dominated by the range $|k_x| < 1$, which is captured by the unstable and stable modes at those wavenumbers to a high precision. The small amount of stresses that are contributed by smaller scales of fluctuations span a broad range of wavenumbers, due to the smaller scales in magnetic fields generated via straining by the flow.} \label{fig:fig13}
\end{figure*}

As can be seen in Fig.~\ref{fig:fig13}, the Reynolds stress is dominated by large scales while the Maxwell stress involves a large number of different scales. Figure~\ref{fig:fig13}(a), using axes with linear scales, shows the dominance of Reynolds stress in the entire system, which the two-eigenmodes-per-wavenumber decomposition (unstable and stable modes) captures, not only qualitatively, but also quantitatively with great accuracy. In Fig.~\ref{fig:fig13}(b), a logarithmic scale is used to expose the range of small scales that contribute significantly to the magnetic fluctuations. Wavenumbers $k_x \lesssim 10$ have major contributions, as opposed to $k_x < 1$ for the fluctuations of the flow. The fact that a large amount of flow energy resides at large scales suggests that the shear-flow turbulence may be amenable to some form of quasilinear modeling. Homogeneous isotropic turbulence, on the other hand, would not be reliably captured with such models, as no scale separation exists therein. Recent studies have highlighted that improved quasilinear models such as the generalized quasilinear approximation are realizable in systems with length- or time-scale separation. \cite{marston2016}

The magnetic fluctuations, on the other hand, span a broad range of scales. This can be physically interpreted as a result of the straining of the magnetic fields by the turbulent flow, which generates small scales in the magnetic fields.\cite{batchelor1950, batchelor1954, townsend1976, schekochihin2002} The straining process by the large-scale turbulent eddies converts the large-scale kinetic energy into the intermediate-scale magnetic energy. \cite{alexakis2005} Magnetic fluctuations at such scales can then, via Lorentz force, feed back on the flow, although mostly at smaller scales. A comprehensive analysis of energy transfer for the present system will be reported in a forthcoming publication where nonlinear mode-coupling and energy transfer between fluctuations of discrete and continuum modes of velocity and magnetic fields are also analyzed.

To model any aspect of magnetic fluctuations, one must thus rely on tools such as statistical theories to obtain scaling laws that can offer insights into these fluctuations. One such approach is detailed next.

%%%%%%%%%%%%%%%%%%%%%%%%%%%%%%%%%%%%%%%%%%%%%%%%%%%%%%%%%%%%%%%%%%%%%%%%%%%%%%%%%%%%%%%%%%%%%%%%%%%%%
%%%%%%%%%%%%%%%%%%%%%%%%%%%%%%%%%%%%%%%%%%%%%%%%%%%%%%%%%%%%%%%%%%%%%%%%%%%%%%%%%%%%%%%%%%%%%%%%%%%%%
\subsection{Scaling law for continuum modes}
Until this point, the discrete modes---unstable and stable modes---which describe the turbulent flow well, have been our focus. The magnetic fluctuations, on the other hand, result from the straining of field lines by the flow, exciting the remaining continuum modes. Hence these modes are necessary for a successful reconstruction of the magnetic fluctuations. Thus, we seek a simple scaling law for the saturated turbulent amplitudes of the continuum modes.

\subsubsection{Analytical prediction for continuum mode amplitudes}
\BT{It is instructive to write the nonlinear MHD equations in the eigenmode basis, arriving at what is also referred to as the mode-amplitude evolution equation\cite{terry2006,fraser2017, tripathi2022, fraser2020thesis, burns2018thesis}
\begin{equation}  \label{eq:threewave}
   \partial_t \beta_j(k_x) = i \omega_j(k_x) \beta_j(k_x) + \sum_{\substack{k_x', k_x'', m,n\\ k_x'+k_x''=k_x}} C_{jmn}(k_x, k_x') \beta_{m}(k_x') \beta_{n}(k_x''),
\end{equation}
where $\beta_j(k_x)$ represents the complex amplitude of an eigenmode $j$ at wavenumber $k_x$ with $\omega_j$ the associated mode-frequency; the nonlinear mode coupling coefficient $C_{jmn}$ measures the three-wave overlap, which dictates the strength of nonlinear beating between an eigenmode $m$ at wavenumber $k_x'$ and an eigenmode $n$ at wavenumber $k_x''$, driving an eigenmode $j$ at wavenumber $k_x$ (with the constraint $k_x'+k_x''=k_x$).}

\BT{For the continuum modes, as was mentioned in Sec.~\ref{sec:sec3a}, their frequencies depend linearly on the wavenumber $k_x$ as \cite{case1960} $\omega/k_x +U_\mathrm{ref}(z) \pm v_{\mathrm{A, ref}}(z)=0$. This implies $\omega_j = \omega  \propto k_x$.}

\BT{Heuristically, the scaling of the nonlinear mode coupling coefficient with wavenumber can be obtained in the following manner: In Eqs.~\eqref{eq:momentumeqn} and \eqref{eq:inductioneqn}, the separation of linear and nonlinear terms arises in Poisson brackets. Consider a prototype equation,
\begin{equation} \label{eq:prototype}
    \begin{aligned}[b]
        \partial_t \widetilde{P} &= \{P,Q \} + ...\\
        &= \{\widetilde{P},\widetilde{Q} \} + \{\widetilde{P}, Q_\mathrm{0} \} + \{P_\mathrm{0}, \widetilde{Q} \} + ...,
    \end{aligned}
\end{equation}
where $P = P_0 + \widetilde{P}$ and $Q = Q_0 + \widetilde{Q}$  represent two fields (e.g., $\nabla^2\phi$ or $\psi$ for the present problem), with $P_0$ representing the $x$-averaged mean component of $P$, and $\widetilde{P}$ standing for perturbations. The linear term, e.g., $i k_x \hat{P} \cdot \partial_z Q_\mathrm{0}$ which is in spectral space, contains only one perturbed field, whereas the nonlinear term, e.g., $i k_x'\hat{P}' \cdot \partial_z \hat{Q}''$, has two perturbed fields, with $\hat{P}'$ and $\hat{Q}''$ representing the Fourier-transformed quantities at wavenumbers $k_x'$ and $k_x''$, respectively. It may be supposed that the derivative $\partial_z$ on the perturbed quantities is roughly on the scale of $|\partial_z| \sim k_x$. (This can be shown analytically for all the eigenmodes, where the background flow is approximately uniform, see Ref.\cite{fraser2017}.) Notice, however, that this argument applies only to the perturbations: the operator $\partial_z$ acting on $Q_0$ clearly does not produce a factor of $k_x$, which is zero for the mean component $Q_0$. We now use this distinction to make a prediction for the amplitudes of perturbations, in particular the continuum mode-amplitudes. The linear and nonlinear terms thus assume the forms $i k_x \hat{P} \cdot \partial_z Q_\mathrm{0}$ and $i k_x'\hat{P}' \cdot i k_x'' \hat{Q}''$, respectively. }

\BT{In Eq.~\ref{eq:prototype}, expanding the perturbations in the eigenmode basis, e.g., $\hat{P} = \sum_l \beta_l \hat{P}_l$ with $\hat{P}_l$ representing the $l$-th eigenmode, and diagonalizing the linear terms (operator), one finds an equation of the form given in Eq.~\ref{eq:threewave}. We can now attempt to understand the behavior of the nonlinear mode coupling coefficients that drive the continuum modes. Assuming nonlinear interactions between the continuum modes are local in spectral space---interaction of three wavenumbers of similar scales---the nonlinear term in Eq.~\ref{eq:prototype} simplifies, e.g.,  $i k_x'\hat{P}' \cdot i k_x'' \hat{Q}''$ becomes $-k_x^2 \hat{P} \hat{Q}$; note the linear term has the form $i k_x \hat{P} \cdot \partial_z Q_\mathrm{0}$. }

\BT{In assuming local interaction between the continuum modes spectral space in $k_x$, the involvement of unstable and stable modes in nonlinear interactions is ignored, which otherwise could bring in non-local effects. This may be a valid assumption for continuum modes at scales much above the Kelvin-Helmholtz-unstable wavenumber range, i.e., $k_x>1$, as the wavenumber convolution constraint of $k_x = k_x' + k_x''$ does not allow two (un-)stable modes to beat together to drive a continuum mode at large $k_x$, e.g., $k_x>2$. }

\BT{Continuing with the above assumption, the nonlinear term $-k_x^2 \hat{P} \hat{Q}$ has one extra $k_x$ compared to the linear term $i k_x \hat{P} \cdot \partial_z Q_\mathrm{0}$. This implies that, for the continuum modes, the nonlinear mode coupling coefficients $C$ are expected to scale as 
\begin{equation}
C \propto k_x^2,
\end{equation}
because the linear term for the continuum modes in Eq.~\ref{eq:threewave} has the eigenfrequency that depends linearly on $k_x$, i.e., 
\begin{equation}
\omega \propto k_x.
\end{equation} 
Such a property of nonlinear coupling coefficient is common in other turbulence calculations, as well. \cite{terry2018}} 

\BT{In order to obtain a phenomenological scaling law, we now make no distinction between different continuum modes, and thus balance the linear and nonlinear terms of Eq.~\ref{eq:threewave} in the quasi-stationary state as $\omega \beta \sim C \beta^2 $. Inserting their asymptotic dependences on $k_x$, the amplitudes of continuum modes is found to follow 
\begin{equation}
\beta \sim k_x^{-1}.
\end{equation} 
Note that the assumptions made in arriving at this simple scaling law are crude. The next step will be to determine from nonlinear simulations whether this scaling can be recovered or whether a number of assumptions made above render the result inapplicable.}

\subsubsection{Numerical verification}

% \begin{figure*}[htbp!]
% 	\includegraphics[width=1\textwidth]{PoP/figures/fig13.png}
% 	\caption{Numerical scaling law of saturated and time-averaged eigenmode amplitudes. Vertical lines show the self-similar cascade along the wavenumber $k_x$. The data presented is for the simulation with $\Ma=10$ and $D_\mathrm{Krook}=2$.}\label{fig:fig13}
% \end{figure*}

\begin{figure}[htbp!]
	\includegraphics[width=1\textwidth]{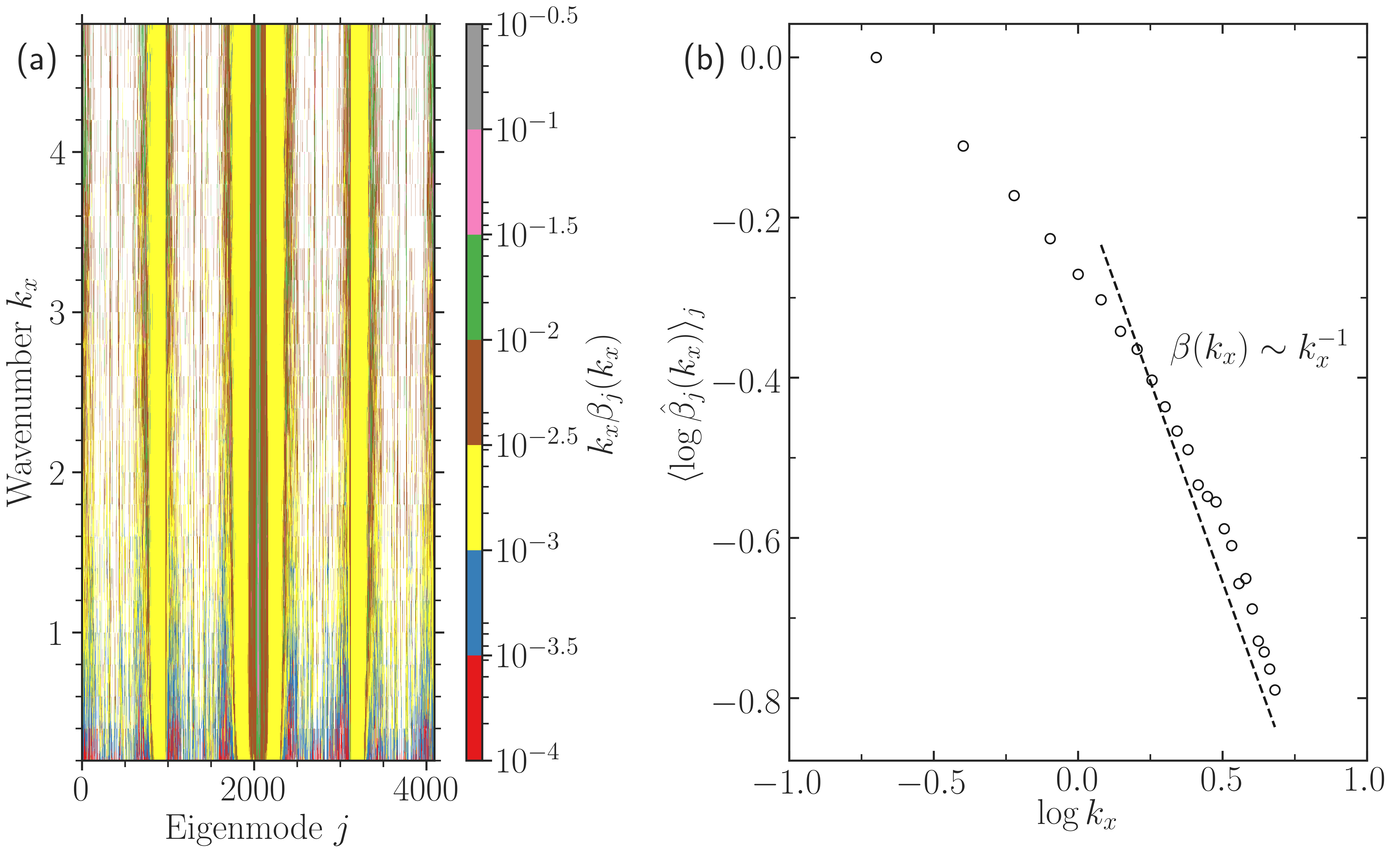}
	\caption{\BT{(a) Dependence of time-averaged eigenmode amplitudes on $k_x$. The indices $j$s of the eigenmodes are arranged in increasing order of their real frequencies. Vertical lines signify the self-similar cascade of energy to small scales in $k_x$. Mode amplitudes are averaged over a quasi-stationary state of turbulence $t=400\text{--}800$ for a simulation with $\Ma=120$ and $D_\mathrm{Krook}=2$. (b) $k_x$ spectra of mode amplitudes in a nonlinear simulation (shown with empty circles) in a $\log_{10}\mathrm{-}\log_{10}$ plot. Shown with a solid line is the analytical prediction, made for the wavenumbers that lie beyond the Kelvin-Helmholtz-unstable range, i.e., for $k_x>1$. The spectral index, predicted based upon a number of simple assumptions, can be seen to fit the data reasonably well.}}\label{fig:fig14}
\end{figure}

Time-averaged eigenmode amplitudes from nonlinear simulations, after multiplying with $k_x$, are plotted in Fig.~\ref{fig:fig14}(a) as functions of $k_x$ and eigenmode index $j$, arranged in order of increasing real frequency of the eigenmodes. The appearance of vertical near-equicontour lines signifies that eigenmodes are excited in a similar pattern across a large range of scales. 

The eigenmodes that lie within the yellow bands are localized in space ($z$-axis), but the band spans a range of heights, outside the shear layer $|z|<1$. Empirically, we note that the center of the rightmost [leftmost] band corresponds to $\omega/k_x = U_{00}$ [$\omega/k_x = -U_{00}$] where $U_{00}=1$. These thick bands represent all eigenmodes that have phase speeds  $\omega/k_x=U_{00} + c v_{\mathrm{A},0}$ [$\omega/k_x=-U_{00} + c v_{\mathrm{A},0}$] where $-1 < c < 1$ and $v_{\mathrm{A},0} = 1/\Ma$; note that $|c|=1$ is not included in these bands. All of these eigenmodes have peaks and oscillations in their eigenfunctions outside of the shear layer. In the layer, the unstable and stable modes maintain their dominance and thus these two discrete modes alone almost completely regulate the momentum transport across the layer, as was noted in Sec.~\ref{sec:transportmodeling}.

It is of interest now to compute from numerical simulation data how the amplitude of each eigenmode $j$ falls off with $k_x$ and construct a $j$-averaged spectral index. To this end, we note the amplitude $\beta_j(k_x=0.2)$ for each mode $j$ at $k_x=0.2$ (the first wavenumber in the simulation) and compute a scaled mode-amplitude $\hat{\beta}_j(k_x)$ as 
\begin{equation}
\hat{\beta}_j(k_x) = \frac{\beta_j(k_x)}{\beta_j{(k_x\mathrm{=}0.2)}},  
\end{equation}
which is expected to fall-off with $k_x$ as $\sim k_x^{\alpha}$. In principle, the spectral index $\alpha$ can depend on the eigenmode index $j$, but a $j$-averaged spectral index is sought now, following the procedure
\begin{equation}
\begin{aligned}[b]
\hat{\beta}_j(k_x) &\propto (k_x/0.2)^\alpha,\\
\mathrm{log\, }\hat{\beta}_j(k_x) &\propto \alpha\, \mathrm{log\, }k_x,\\
\langle \mathrm{log\, }\hat{\beta}_j(k_x)\rangle_j &\propto \langle \alpha \rangle_j\, \mathrm{log\, }k_x.\\
\end{aligned}
\end{equation}
This $j$-averaged spectral distribution of the mode-amplitudes informs how, on average, each eigenmode amplitude depend on $k_x$.

A plot of $\langle \mathrm{log\, }\hat{\beta}_j(k_x)\rangle_j$ vs. $\mathrm{log\, }k_x$ is shown in Fig.~\ref{fig:fig14}(b), along with the analytical prediction of inverse-in-wavenumber fall-off of the mode amplitudes, at scales above the Kelvin-Helmholtz-unstable wavenumber range. It should be highlighted that the computation of all the eigenmode amplitudes at each wavenumber at each simulation time is computationally demanding, as the process requires the computation of modified left eigenmodes for each right eigenmode at each wavenumber, apart from the mode projection calculation at each time step. Therefore, only the first $24$ Fourier modes in $k_x$ are shown in Fig.~\ref{fig:fig14}.

A finding in Fig.~\ref{fig:fig14} is the identification of self-similar cascade of energy to smaller scales (larger $k_x$) in \textit{eigenmode space}. This result also hints that the interaction involving the continuum modes may be reasonably simplified, and potentially be valuable in estimating the amplitudes of unstable and stable modes, using closure theories (see Ref.\cite{terry2018} for a recent example).

\section{Conclusions}\label{sec:sec6}

We have investigated MHD turbulence in two-dimensions, driven by a forced unstable shear flow, using a complete eigenmode decomposition of fluctuations in nonlinear simulations, which exposes the nonlinearly saturated excitation level of each eigenmode and its role. Intrinsic to linear instability, the unstable modes derive fluctuation energy from the mean flow gradient. The linearly-decaying stable modes, however, contain almost the same amount of energy as the unstable mode, which they receive via nonlinear excitation. This truncated basis of two eigenmodes per wavenumber is found to reconstruct essential large-scale features of turbulent flow and the associated momentum transport via Reynolds stress. Quantifying transport due to unstable modes alone shows an overestimation up to an order of magnitude higher relative to the actual flux. The reduction in the flux is identified to be due to the continuous up-gradient transport by the stable modes, which causes a near-cancellation of down-gradient transport driven by unstable modes.

The continuum modes, on the other hand, describe small-scale fluctuations of the flow and magnetic field, where the above large-scale unstable and stable modes manifest themselves as a quasi-coherent vortex. To predict the mode amplitudes of the continuum, a simple scaling law is derived from the governing nonlinear equations and the predicted inverse-in-wavenumber fall-off rate is found reasonably agree with the simulation data.

Although both the momentum transport and fluctuation energy are largely described by the discrete modes, the former is more efficiently captured [Figs.~\ref{fig:fig9}\textrm{--}\ref{fig:fig13}] as almost all the momentum transport occurs near $z=0$, which is the region where the discrete modes dominate [Figs.~\ref{fig:fig1}(b)~and~\ref{fig:fig1}(c)]. The fluctuation energy, on the other hand, is related to fluctuations that are scattered in and around the large-scale eddies; a portion of this energy is claimed by the continuum modes, although a large fraction still belongs to the discrete modes [Figs.~\ref{fig:fig5}(b),~\ref{fig:fig6}\textrm{--}\ref{fig:fig8}].

Transport reduction by stable modes can also be used to improve phenomenological constructs like eddy viscosity, which are generally agnostic as to the nonlinear excitation of stable modes. By predicting the turbulent amplitudes of the unstable and stable modes, for astrophysically relevant parameters, e.g., very large $\Rm$, $\Re$, $\Pm$ compared to unity, the simple relation between turbulent viscosity and eigenmode amplitudes [in Eq.\eqref{eq:turbvisc}] can be exploited to reliably model transport processes in astrophysical objects, which otherwise cannot be solved using current state-of-the-art direct numerical simulations. It should be noted that such a prediction for the mode amplitudes $|\beta_1|$ and $|\beta_2|$ was recently made for ion-temperature-gradient-driven fusion plasma turbulence \cite{terry2021} using statistical closure theory. \cite{orszag1970} Undertaking such a task for the present system is interesting, but beyond the purview of this work and will thus be left for future investigations.

The reduced representation of turbulent flow and transport presented here is also useful for building sub-grid-scale models, which can allow performing nonlinear simulations at extreme parameters with less-intensive computational demands. Progress can thus be made in seeking models that reduce the number of degrees of freedom while capturing essential features of the turbulent system. Techniques like proper orthogonal decomposition, dynamic mode decomposition, etc., also exist for such purposes, \cite{taira2017} but they operate on output from nonlinear simulations, and it can be difficult to assign intuitive physical meaning to the characteristic mode structures. Here, the truncated eigenmode basis, composed of the unstable and stable modes, has been demonstrated to reconstruct nonlinear fluctuations to an appreciable degree, thus suggesting that they can be leveraged as a physically-motivated basis for extreme parameter studies, without having to first perform a direct numerical simulation. These modes may also be useful for generating, via their nonlinear interactions with continuum modes, fluctuations associated with the continuum modes. Such a test could be performed to analyze magnetic fluctuations. The reduced basis, composed of unstable and stable mode alone, can also serve in direct statistical simulations, \cite{allawala2020} which have shown promises towards simulating the slowly-evolving turbulent statistics, e.g., two-point two-time correlations, three-point correlations between fluctuating fields, etc., rather than the fast-evolving field variables themselves, e.g., flow velocities. Other improved forms of quasilinear models like the generalized quasilinear approximation\cite{marston2016} may also benefit from using this truncated basis. This possibility will be explored in a separate publication.

In the future, procedures similar to that employed here can be used to examine the properties of other forms of instability-driven turbulence, such as magneto-rotational-instability-driven \cite{pessah2006} and stratified-shear-flow-driven turbulence.\cite{garaud2018} Building nonlinear energy transfer diagnostics in shear-flow turbulence to study the physical processes and scales that impact the difference in unstable- and stable-mode amplitudes is another possible avenue. Such investigations constitute steps towards deployment in service to one-dimensional stellar transport models. \cite{fuller2019} Central to improved predictiveness are the stable modes, whose properties will similarly require additional studies.

%%%%%%%%%%%%%%%%%%%%%%%%%%%%%%%%%%%%%%%%%%%%%%%%%%%%%%%%%%%%%%
%%%%%%%%%%%%%%%%%%%%%%%%%%%%%%%%%%%%%%%%%%%%%%%%%%%%%%%%%%%%%%%%%%%%%%%%%%%%%%%%%%%%%%%%%%%%%%%%%%%%%

\begin{acknowledgments}
The authors appreciate K.~Burns, E.H.~Anders, and the Dedalus developers for assistance in several aspects of leveraging the numerical code. Thanks also to J.~Fuller and other participants of the program ``Transport in Stellar Interiors, 2021" at the Kavli Institute of Theoretical Physics for useful discussions. This material is based upon work funded by the Department of Energy [DE-SC0022257] through the NSF/DOE Partnership in Basic Plasma Science and Engineering. We also gratefully acknowledge support from NSF Grant Nos.~AST-1814327 and AST-1908338. The simulations reported herein were performed using the XSEDE supercomputing resources via Allocation No.~TG-PHY130027. 

The data that support the findings of this study are available from the corresponding author upon reasonable request.
\end{acknowledgments}

\section*{Appendix A: Orthogonality of right eigenmode and modified left eigenmode}\label{sec:AppendixA}

%%%%%%%%%% Prefix a "S" to all equations, figures, tables and reset the counter %%%%%%%%%%
\setcounter{equation}{0}
\setcounter{figure}{0}
\setcounter{table}{0}
\makeatletter
\renewcommand{\theequation}{A\arabic{equation}}
\renewcommand{\thefigure}{A\arabic{figure}}

\BT{Due to the non-normality of the linear operator of the shear-flow instability, the eigenmodes are not orthogonal. This presents a significant challenge in the computation of mode amplitudes. An additional challenge is the generalized eigenvalue nature of the problem at hand, when written in vorticity formalism, as in Eq.~\eqref{eq:lineqnsa}. This differs from the standard eigenvalue problem, $L \xi_j=\lambda_j \xi_j$, where $L$ is a linear operator whose $j$-th eigenmode is $\xi_j$ with eigenvalue $\lambda_j$. The generalized eigenvalue problem that we encounter here is
\begin{equation} \label{eq:appendix1}
    L X_j=\omega_j M X_j,
\end{equation}
where $L$ is a linear operator, $M$ is another linear operator containing the Laplacian operation for our problem, $X_j$ is the $j$-th (right) eigenmode with corresponding (right) eigenvalue $\omega_j$. [Often times, the distinction between left and right eigenmodes of a linear operator is not made as they happen to be the same; however this is not the case here for the non-normal operator.] The right eigenmodes, although non-orthogonal to each other, can be made orthogonal with an appropriate weight factor to the left eigenmodes, which are solutions to another eigenvalue problem: \cite{burns2018thesis, fraser2020thesis}
\begin{equation} \label{eq:appendix2}
    Y_j^\mathrm{T} L=\sigma_j Y_j^\mathrm{T} M.
\end{equation}
Here, $Y_j^{\mathrm{T}}$ is the transpose of the left eigenmode with its left eigenvalue $\sigma_j$. A slight reformulation is possible to this equation by taking Hermitian-transpose:
\begin{equation}
    L^\dagger Y_j^\ast=\sigma_j^\ast  M^\dagger Y_j^\ast.
\end{equation}
In the eigenvalue solver in Dedalus, the matrices $L^\dagger$ and $M^\dagger$ for each wavenumber are passed, and their eigenmodes $Y_j^\ast$ and eigenvalues $\sigma_j^\ast$ are found. It can be shown that the eigenvalues $\sigma_j$ and $\omega_j$ are the same (i.e., $\sigma_j = \omega_j$), by analyzing Eqs.~\eqref{eq:appendix1} and \eqref{eq:appendix2}. A modified orthogonality relation between the left and right eigenmodes can now be derived: \cite{burns2018thesis, fraser2020thesis}
\begin{equation}
\begin{aligned}[b]
    &(Y_j^\mathrm{T} L) X_i = Y_j^\mathrm{T} (L X_i) = Y_j^\mathrm{T} (\omega_i M X_i)\\
    \implies &(\sigma_j Y_j^\mathrm{T} M) X_i= Y_j^\mathrm{T} (\omega_i M X_i)\\
    \implies &Y_j^\mathrm{T} M X_i (\sigma_j-\omega_i) =0\\
    \implies &Y_j^\mathrm{T} M X_i \propto \delta_{i,j},
\end{aligned}
\end{equation}
which means the left and right eigenmodes are orthogonal to each other with a weight factor $M$, as long as their eigenvalues differ ($\sigma_j\neq \omega_i$). For numerical computation, it is convenient to define $Y_j^\mathrm{T} M = (M^\mathrm{T} Y_j)^\mathrm{T} = Z_j^\mathrm{T}$ where $Z_j$ is the \textit{modified} left eigenmode, which is---by construct---orthogonal to the right eigenmode without any weight factor: $Z_j^\mathrm{T} X_i \propto \delta_{i,j}$. Using this relation the eigenmode coefficients $\beta_j({k_x,t})$ in the eigenmode expansion of turbulent fluctuations are computed at each wavenumber and at each time.}

\section*{Appendix B: Cyclic oscillations in mode-amplitudes for weak magnetic fields}\label{sec:AppendixB}

%%%%%%%%%% Prefix a "S" to all equations, figures, tables and reset the counter %%%%%%%%%%
\setcounter{equation}{0}
\setcounter{figure}{0}
\setcounter{table}{0}
\makeatletter
\renewcommand{\theequation}{B\arabic{equation}}
\renewcommand{\thefigure}{B\arabic{figure}}

\begin{figure*}[htbp!]
	\includegraphics[width=1\textwidth]{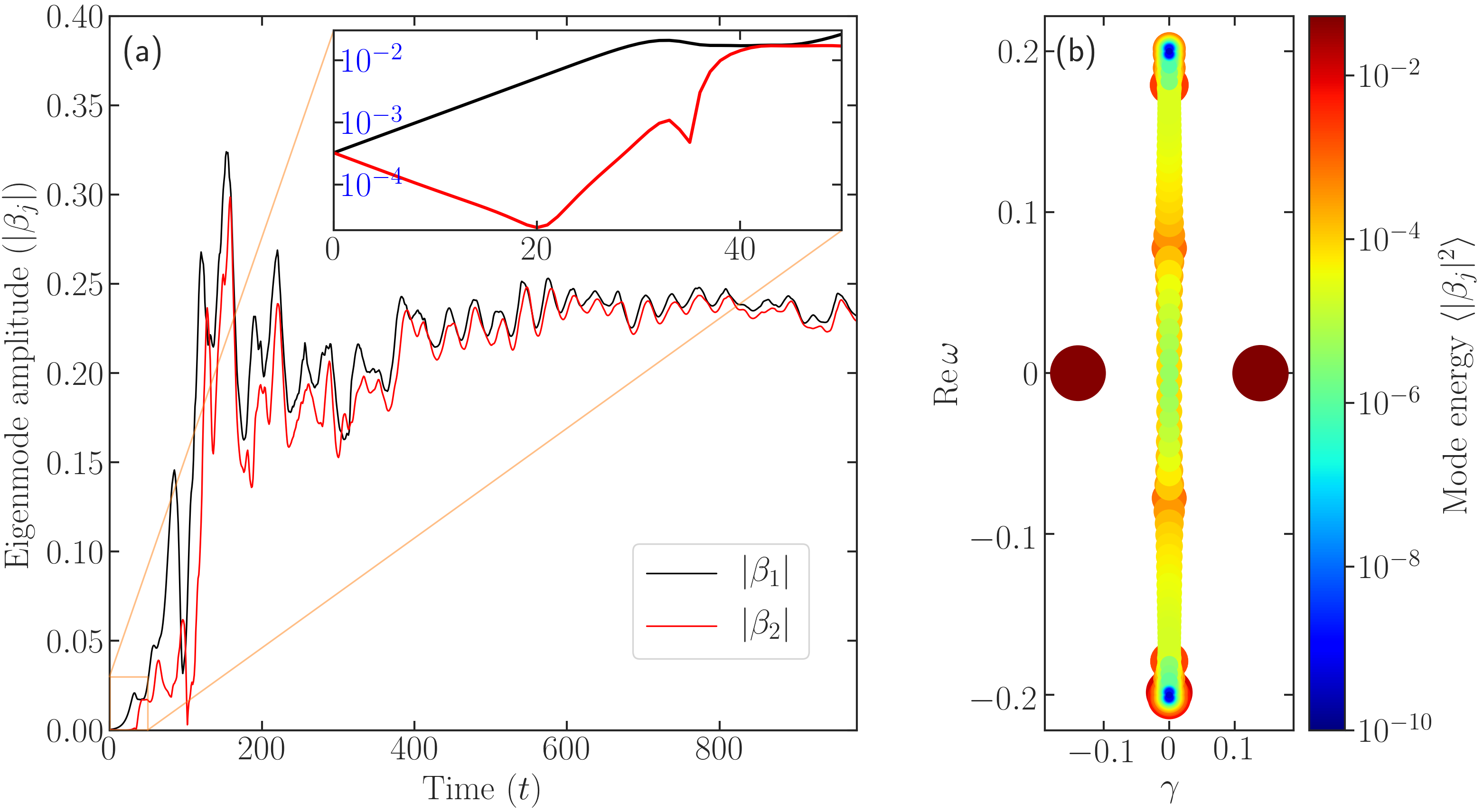}
	\caption{Same as in Fig.~\ref{fig:fig5}, but for $\Ma=120$. The amplitude of the nonlinearly excited stable mode is almost exactly the same as that of the unstable mode. The nature of their oscillations are also similar, although the oscillations in the stable-mode-amplitude lags behind that of the unstable mode. The lag is likely an outcome of a time-delay in energy transfer from the unstable to the stable mode at the same wavenumber, which thus requires a series of nonlinear interactions with fluctuations at other wavenumbers.}\label{fig:fig15}
\end{figure*}

\end{document}